\documentclass[preprint,12pt]{elsarticle}
\usepackage{amssymb}

\journal{Astroparticle Physics}

\newcommand{\be}{\begin{equation}}
\newcommand{\ee}{\end{equation}}
\newcommand{\simless}{\lower.5ex\hbox{$\; \buildrel < \over \sim\;$}}
\newcommand{\simgreat}{\lower.5ex\hbox{$\; \buildrel > \over \sim\;$}} 
 
\newcommand{\mbar}{{ \langle m \rangle }} 
\newcommand{\rhocen}{{ \rho_{\rm c}}} 
\newcommand{\thetacen}{{ \Theta_{\rm c} }} 

\newcommand{\tcent}{{ T_{\rm c} }}
\newcommand{\tig}{{ T_{\rm nuc} }}

\newcommand{\rgas}{{ {\cal R} }} 
\newcommand{\mzero}{{ \mu_0 }} 
\newcommand{\conlum}{{ {\cal C} }} 
\newcommand{\yield}{{ \langle \Delta E \rangle }}

\newcommand{\bcon}{{ f_{\rm g} }}

\newcommand{\mpro}{ m_{\rm p}} 
\newcommand{\starmass}{ M_0 } 

\newcommand{\econ}{ {\rm e} } 
\newcommand{\aabe}{\alpha(\alpha,\gamma)} 

\begin{document}

\begin{frontmatter}

\title{{\bf Stellar Helium Burning in Other Universes:} \\
A Solution to the Triple Alpha Fine-Tuning Problem} 

\author[a,b]{Fred C. Adams}
\author[a]{and Evan Grohs}

\address[a]{Physics Department, University of Michigan, Ann Arbor, MI 48109} 
\address[b]{Astronomy Department, University of Michigan, Ann Arbor, MI 48109} 


\begin{abstract}
Motivated by the possible existence of other universes, with different
values for the fundamental constants, this paper considers stellar
models in universes where $^8$Be is stable.  Many previous authors
have noted that stars in our universe would have difficulty producing
carbon and other heavy elements in the absence of the well-known
$^{12}$C resonance at 7.6 MeV. This resonance is necessary because
$^8$Be is unstable in our universe, so that carbon must be produced
via the triple alpha reaction to achieve the requisite abundance.
Although a moderate change in the energy of the resonance (200 -- 300
keV) will indeed affect carbon production, an even smaller change in
the binding energy of beryllium ($\sim100$ keV) would allow $^8$Be to
be stable. A stable isotope with $A=8$ would obviate the need for the
triple alpha process in general, and the $^{12}$C resonance in
particular, for carbon production. This paper explores the possibility
that $^8$Be can be stable in other universes. Simple nuclear
considerations indicate that bound states can be realized, with
binding energy $\sim0.1-1$ MeV, if the fundamental constants vary by a
$\sim {\rm few}-10$ percent. In such cases, $^8$Be can be synthesized
through helium burning, and $^{12}$C can be produced later through
nuclear burning of beryllium. This paper focuses on stellar models
that burn helium into beryllium; once the universe in question has a
supply of stable beryllium, carbon production can take place during
subsequent evolution in the same star or in later stellar
generations. Using both a semi-analytic stellar structure model as
well as a state-of-the-art stellar evolution code, we find that viable
stellar configurations that produce beryllium exist over a wide range
of parameter space. Finally, we demonstrate that carbon can be
produced during later evolutionary stages.
\end{abstract}  


\end{frontmatter}

\section{Introduction} 
\label{sec:intro} 

Although carbon is a crucial element for the existence of life in our
universe, its production in stars is not straightforward.  Carbon must
be produced via the triple alpha reaction (see below), and the
delicate nature of this process has been often used as an example of
the so-called fine-tuning of our universe for life
\cite{carr,bartip,hogan,aguirre,barnes2012}.  This paper offers an
alternate solution to this apparent fine-tuning problem. If the
fundamental constants of nature are sufficiently different, so that
the triple alpha reaction is compromised, then a large fraction of
parameter space allows for stable $^8$Be nuclei, which obviates the
need for the triple alpha reaction. In such universes, helium burning
leads to (stable) beryllium, and subsequent beryllium burning can then
produce carbon, oxygen, and other large nuclei. As long as stable
$^8$Be exists, all of these nuclei can be synthesized through
non-resonant two-body reactions, with no need for the triple alpha
process or any particular resonance.

The necessity of the triple reaction for carbon production
\cite{clayton,kippenhahn,hansen} can be summarized as follows. The key
issue is that our universe supports no stable nuclei with atomic mass
number $A=8$. For a universe starting with a composition of primarily
hydrogen and helium, the $^8$Be nucleus is the natural stepping stone
towards carbon; unfortunately, it decays back into its constituent
alpha particles with a half-life of only about $\sim10^{-16}$ sec. As
a result, the synthesis of carbon depends on the triple alpha process 
\be
3 \alpha \to ~^{12}{\rm C} + \gamma \,, 
\ee
where three helium nuclei combine to make carbon. This reaction, 
in turn, relies on the temporary formation of $^{8}$Be nuclei
\cite{salpeter}. In spite of the instability of $^8$Be, a small and
transient population of these nuclei builds up and allows for the
fusion of a third alpha particle.  In order for the reaction to take
place fast enough, however, the final step ($^8$Be + $\alpha \to$
$^{12}$C) must take place in a resonant manner. The existence of this
resonance was famously predicted by Hoyle \cite{hoyle} and the
corresponding energy level in carbon was subsequently measured in the
laboratory (e.g., see the review of \cite{fowler}).

Stellar nucleosyntheis calculations \cite{clayton,kippenhahn,hansen}
indicate that the triple alpha process is necessary to produce the
observed abundance of carbon in our universe. Subsequent work has
shown that the energy level of the resonance, which occurs at 7.644
MeV, cannot vary by a large increment without compromising carbon
production.  As one example, calculations of helium burning in 20
$M_\odot$ stars have been carried out \cite{livio} using different
values for the location of the resonance --- but continuing to require
the triple alpha reaction. These results show that a 60 keV increase
in the crucial energy level in the $^{12}$C nucleus does not
significantly alter carbon production in stellar interiors. In
contrast, however, an increase of 277 keV (or more) in the energy
level leads to different nuclear evolution and much less carbon is
synthesized (most of the carbon is burned into oxygen). More recent
calculations \cite{eckstrom,epelbaum,oberhummer,schlattl} reach
similar conclusions, although the amount of carbon produced depends on
both the mass of the star and the stellar evolution code (e.g.,
compare results from \cite{livio}, \cite{oberhummer}, and
\cite{schlattl}; see also \cite{higa} for further discussion of
nuclear structure and \cite{schellekens} for a more comprehensive
review).

The above considerations indicate that the resonant energy for
$^{12}$C can only change by an increment of order 100 keV without
compromising carbon production via the triple alpha process. We note
that {\it some} carbon can be produced without the triple alpha
resonance, but the abundance would be much lower than that of our
universe; since we do not know the minimum carbon abundance that is
necessary for life, it is not known if such low carbon universes would
be habitable. In addition, the triple alpha process is only necessary
because $^8$Be is unstable, and its binding energy is higher than two
separate alpha particles by a difference of about 92 keV. The
similarity of these two energy scales suggests the following solution
to the triple alpha fine-tuning problem. If the $^{12}$C resonance
changes enough to alter carbon production via the triple alpha
process, then $^8$Be will often be stable, so that the triple alpha
reaction is no longer necessary. Of course, the changes in nuclear
structure must result in the binding energy of $^8$Be being larger 
(more bound), rather than lower, so that only ``half'' of the
parameter space would be viable.

The scenario considered here implicitly assumes that big bang
nucleosynthesis proceeds in essentially the same manner as in our
universe, with comparable light element abundances. The exact
abundances are expected to vary, of course, as we assume here that
nuclear physics changes enough to allow for stable $^8$Be. In our
universe, $^7$Li is stable and can be more easily produced than
heavier isotopes, but the abundance is only $\sim10^{-10}$. The 
abundance of stable $^8$Be in this alternate scenario is expected to
be even lower.  We thus assume that the universe emerges from its
early epochs with a composition dominated by hydrogen and helium, and
relatively little mass in heavier isotopes.

For universes in this regime of parameter space, the key requirement
for carbon production is that helium burning can take place readily to
form $^8$Be. As long as the universe under consideration can produce
$^8$Be, and it retains an appreciable supply of $^4$He, carbon can be
produced in subsequent stellar generations through the reaction $^4$He
+ $^8$Be $\to$ $^{12}$C. This reaction is a natural channel to produce
carbon --- alpha particles are energetically favorable states, so that
isotopes produced by adding together alpha particles would naively be
the easiest to make. Note that in our universe the most common
isotopes (after $^1$H) are $^4$He, $^{16}$O, $^{12}$C, and $^{20}$Ne,
in decreasing order of abundance. In addition, the isotopes $^{28}$Si,
$^{24}$Mg, and $^{32}$S are also among the top ten most common. All of
the small alpha-particle nuclei are thus well-represented with the
exception of $^8$Be.

An underlying assumption of this paper is that the fundamental
constants of nature can take on varying values. For completeness, we
note that this possibility arises in two separate but related
contexts. Within our universe, the constants of nature could vary with
time, although observations limit such variations to be quite small
\cite{barrow,davies,uzan}.  On a larger scale, the constants of nature
could vary from region to region within the vast complex of universes
known as the multiverse \cite{reessix,vilenkin}. This latter
possibility is often invoked as a partial explanation for why the
constants have their observed values. Within the larger ensemble, each
constituent universe samples the values of its constants from some
underlying distribution. Our local region of space-time --- our
universe --- thus has one particular realization of these parameters.
Finally, it is useful to consider different values for the constants 
of nature, independent of the multiverse, in order to understand the 
sensitivity of stellar structure to the relevant input parameters. 

The number and choice of parameters needed to specify a given universe
is relatively large, but no general consensus exists regarding the
relevant parameter space \cite{hogan,dirac,tegmark}. For example,
Table 1 of Ref. \cite{tegmark} includes 37 parameters, whereas
Ref. \cite{reessix} includes only 6.  The range of allowed variations
in the constants also remains undetermined. In previous work, we have
considered how stars are influenced by variations in the fine
structure constant and the gravitational constant, as well as nuclear
parameters determined by the strong and weak forces 
\cite{adams,adamsnew,coppess}. The first two quantities are specified
by the dimensionless parameters 
\be 
\alpha = {e^2 \over \hbar c}
\qquad {\rm and} \qquad 
\alpha_G = {G \mpro^2 \over \hbar c} \,,
\label{alphas} 
\ee  
where $\mpro$ is the proton mass. For the realization of these
constants found in our universe, these parameters have the values
$\alpha\approx1/137$ and $\alpha_G \approx 5.91 \times 10^{-39}$.
This past work (see also \cite{barnes2016}) shows that the fundamental
constants can vary by several orders of magnitude and still allow for
functioning stars and habitable planets. In this work, we keep the
($\alpha$,$\alpha_G$) fixed and explore the implications of different
nuclear structures such as stable $^8$Be.

The solution to the triple alpha fine-tuning problem considered here
requires two components. First, relatively small changes to the
fundamental constants must allow for $^8$Be nuclei that are stable and
hence have lower binding energy than two alpha particles.  Second,
stars in such universes must be able to burn helium into stable
beryllium, and then later burn the beryllium into carbon (and heavier
elements). This paper considers nuclear models to address the first
issue and stellar structure models to address the second. For purposes
of showing that stable bound states of $^8$Be exist, the nuclear
potential is the key quantity and its depth is specified by the strong
coupling constant. Here we show that relatively small increases in the
strong force lead to a stable bound state. As long as a stable state
exists, the behavior of $^8$Be in stellar interiors is determined by
composite parameters $\conlum$ \cite{adams,adamsnew,barnes2016} that 
specify the rates and yields for a given nuclear reaction (see Section
\ref{sec:models}). Note that the allowed nuclear structures, and hence
the values of $\conlum$, are complicated functions of the more
fundamental parameters appearing in the Standard Model of particle
physics (compare Refs.
\cite{hogan,aguirre,livio,epelbaum,barrow,adams,jeltema}).  Using a
semi-analytic stellar structure model \cite{adams}, Section
\ref{sec:models} shows that a large fraction of parameter space allows
for successful helium burning into beryllium, provided that $^8$Be is
a stable isotope. These results are then verified using the
state-of-the-art stellar structure code {\sl\small MESA}
\cite{paxton,paxtonb} in Section \ref{sec:mesa}. Although the entire
parameter space of possible nuclear reactions is too large to fully
explore in this work, we demonstrate that stars can successfully burn
beryllium into carbon in later stages of evolution (a full exploration
of this parameter space is left for future work).  The paper concludes
in Section \ref{sec:conclude} with a summary of our results and a
discussion of their implications.

\section{Nuclear Considerations for Beryllium Bound States} 
\label{sec:nuke} 

This section considers models for nuclear structure in order to
elucidate the requirements for making stable $^8$Be nuclei. Given the
difficulties inherent in making {\it a priori} models for nuclear
structure, along with the large parameter space available for
alternate universes, we consider several approximate approaches.
First, we review previous calculations using lattice chiral effective
field theory \cite{epelbaum}. These results indicate that shifts in
the fundamental parameters at the level of $\sim2-3\%$ are large
enough to allow for stable $^8$Be (Section \ref{sec:lattice}). Next,
for comparison, we use simple approaches where the $^8$Be nucleus is
considered as a bound state of two alpha particles \cite{hafteller}.
Under this assumption, we model beryllium bound states using a square
well potential (Section \ref{sec:square}) and a generalization of Bohr
theory (Section \ref{sec:bohr}). For both of these approximations, the
$^8$Be nucleus can have a stable configuration with binding energy
$\sim$1 MeV lower than separate alpha particles, provided that the
strong coupling constant is larger by a few percent. For completeness,
we also present an order of magnitude estimate using the semiempirical
mass formula, which is based on the liquid drop model (Section
\ref{sec:semf}). This latter argument also indicates that an increase
in the strong coupling constant of a few percent would be sufficient
to provide a stable $^8$Be nucleus. Although approximate, all four of
these approaches suggest that moderate changes in the fundamental
constants would allow for a stable $^8$Be isotope.

\subsection{Lattice Chiral Effective Field Theory} 
\label{sec:lattice} 

This section reviews nuclear models resulting from Lattice Chiral
Effective Field Theory \cite{epelbaum} and uses the results to explore
the possibility of finding bound states for $^8$Be.  The key quantity
of interest is the energy difference between a bound state of $^8$Be
and two separate alpha particles. This energy difference can be
written in the form 
\be
\Delta E_b = E_8 - 2 E_4\,,
\ee 
where $E_8$ and $E_4$ denote the binding energies of $^8$Be and
$^4$He. In our universe, $\Delta E_b \approx +92$ keV, whereas we
require $\Delta E_b<0$ for the synthesis of $^8$Be to be energetically
favored.  The required change is thus of order 100 keV.

The existing calculations for lattice Effective Field Theory
treatments of nuclear structure \cite{epelbaum} consider possible
changes in binding energy due to variations in the pion mass $M_\pi$
and the fine structure constant $\alpha$. The changes in binding
energies for the nuclei of interest can be written in the form 
\be
\delta E_j = {\partial E_j \over \partial M_\pi}\Bigg|_0
\delta M_\pi + {\partial E_j \over \partial \alpha}\Bigg|_0
\delta \alpha \,,
\label{ederiv} 
\ee
where the subscripts on the energies label the nuclear species and
where the partial derivatives are to be evaluated at the values
realized within our universe. This expression represents the leading
order correction and is limited to small changes in the binding
energies. In this context, the total binding energy of $^8$Be (or
twice that of $^4$He) is of order 56 MeV, whereas we are interested in
changes of order 0.1 -- 1 MeV, so that $(\delta E_j)/E_j\ll1$ is 
satisfied. More specifically, this deriviation is based on reference 
\cite{epelbaum}, which holds for relative changes as large as 
$\sim10\%$. In this scheme, the pion mass is simply related to 
the light quark masses $M_\pi^2 \propto m_u + m_d$ \cite{epelbaum}, 
so that the first term in equation (\ref{ederiv}) specifies the
dependence on $(m_u,m_d)$. The second term arises from variations 
in the fine structure constant $\alpha$. 

In order for variations in the constants of nature to allow for stable
$^8$Be nuclei, we require that the binding energy of beryllium change
by an amount that is not equal to twice the binding energy of helium,
i.e., we require $\delta E_8 \ne 2 (\delta E_4)$. This condition can
be realized as long as the derivatives from equation (\ref{ederiv}) do
not have specific values, i.e., we required $\partial E_8/\partial
M_\pi$ $\ne$ 2 $\partial E_4/\partial M_\pi$ and/or $\partial
E_8/\partial \alpha$ $\ne$ 2 $\partial E_4/\partial \alpha$.
Reference \cite{epelbaum} carries out a calculation of these partial
derivatives using an Auxiliary Field Quantum Monte Carlo (AFQMC)
scheme. In the calculation, the derivatives appearing in equation
(\ref{ederiv}) are expanded into separate terms corresponding to
variations in nuclear parameters appearing in the AFQMC action
(strength of the pion exchange potential, the coefficients of the
isospin terms, etc.). Significantly, for the two nuclear species of
interest, the individual partial derivatives (and hence their sums) 
always satisfy the requirement that $\partial E_8/\partial\xi$ $\ne$
$2\partial E_4/\partial\xi$, where $\xi$ represents any of the 
aforementioned parameters. As a result, changes in the nuclear 
parameters (and $\alpha$) can lead to variations in the binding 
energy of $^8$Be. 

The energy difference $\Delta E_b$ can be made either larger or
smaller through variations in the fundamental constants. As a result,
only about half of the allowed variations are expected to lead to
nuclear states with lower binding energies. In addition, a bound state
for $^8$Be requires that the binding energy be lower by at least
$(\Delta E)/E\sim100$ keV/(56 MeV) $\sim0.0018$. This level of
variation in the binding energy of $^8$Be corresponds to changes in
the light quark masses of $2-3\%$ and/or changes of $\sim2.5\%$ in the
fine structure constant \cite{epelbaum}.

\subsection{Square Well Potential} 
\label{sec:square} 

Here we assume that the alpha particles are the basic constituents,
and use a square well potential to model the bound state of the two
particle system (see \cite{davies} for an analogous treatment of
diprotons). The reduced mass is given by $m_R$ = $m_1 m_2/(m_1+m_2)$ 
= $m_\alpha/2\approx2 m_N$, where $m_N$ is the nucleon mass (assumed
to be the same for protons and neutrons). The system dynamics is then
given by the Schr{\"o}dinger equation in its usual form.  Let $E$ be
the energy of the two particle system, which can be bound or unbound,
but does not include the binding energy of the alpha particles
themselves.  Here we assume that the potential can be described by a
three dimensional square well of depth $V_0$ and width $b$ and limit
the discussion to spherically symmetric states ($\ell$ = 0).  For
bound states, the energy $E<0$, so we define $\epsilon \equiv - E =
|E|$, as well as the ancillary quantities

\be
\omega^2 \equiv {2m_R (V_0 - \epsilon) \over \hbar^2} 
\qquad {\rm and} \qquad k^2 \equiv {2m_R \epsilon \over \hbar^2} \,.
\ee
The energy levels are given by the quantum condition 
\be
\omega {\cos \omega b \over \sin \omega b} = - k \,. 
\ee
For convenience we define
\be
z^2 = k^2 b^2 \qquad {\rm and} \qquad 
\lambda^2 = {2 m_R V_0 b^2 \over \hbar^2} \,,
\ee
so that the quantum condition reads 
\be
(\lambda^2 - z^2)^{1/2} \cos (\lambda^2 - z^2)^{1/2} = 
- z \sin (\lambda^2 - z^2)^{1/2} \,.
\ee
We are interested in the case where $z^2 \ll \lambda^2$. In the limit
$z\to0$, we obtain the solution $\lambda_0 = \pi/2$.  This value
represents the critical case, so that no bound states exist for
$\lambda < \lambda_0 = \pi/2$.  In other words, in order for a bound
state to exist, we require $\lambda> \pi/2$, or
\be
V_0 b^2 > {\pi^2 \hbar^2 \over 8 m_R} \,. 
\ee
The right hand side has numerical value $\sim 25.5$ MeV fm$^2$. 

For small (but nonzero) values of $z$, we get 
\be
\lambda \cos \lambda = -z \sin \lambda + {\cal O}(z^2) \,. 
\ee
In this regime, $\lambda$ will be near (but not equal to) $\pi/2$, so
we write $\lambda = \pi/2 + \theta$, where $|\theta|\ll1$, so that
$\cos\lambda=-\sin\theta$ and $\sin\lambda=\cos\theta$.  Now the
quantum condition takes the form 
\be
\left( {\pi \over 2} + \theta \right) \sin\theta = z \cos\theta\,,
\ee
which implies that $z\approx(\pi/2)\theta$, where we have used the
fact that $\theta^2 \ll 1$.  The energy level becomes 
\be
\epsilon = {\pi^2 \hbar^2 \over 8m_R b^2} 
\left[ \left( {2m_R V_0 b^2 \over \hbar^2} \right)^{1/2} 
- {\pi \over 2} \right]^2 \,. 
\ee

Now we want to use this result to estimate out how much change is
required for $^8$Be to have a bound state.  In our universe, with no
bound state, the effective depth $V_0$ of the potential well is close
to the boundary given by $\lambda = \pi/2$. In an alternate universe, 
suppose that the potential well has depth 
\be 
V = f V_0 \,,
\ee
where $f>1$ is a dimensionless factor. For the sake of definiteness, 
we set $b$ = 1 fm. Inserting numerical values, the energy of the 
bound state is given by
\be
\epsilon = 62.7 \, {\rm MeV} \, 
\left( \sqrt{f} - 1 \right)^2 \,. 
\ee
Suppose, for example, we require the bound state to have binding energy
$\epsilon$ = 1 MeV. The required factor $f \approx 1.27$.  If the
binding energy is only 0.1 MeV, the required factor $f \approx$
1.081. As a result, working within the square well approximation, the
depth of the potential well $V_0$ must change by of order a few
percent to allow for a bound state of $^8$Be.\footnote{Since $f=1$
  gives $\epsilon=0$ in this expression, we are considering the actual
  binding energy of $-92$ keV for $^8$Be in our universe to be
  negligible relative to 62.7 MeV.}

Of course, the direction of the change matters. The deeper value of
the potential well arises from an increase in the strength of the
strong force. A decrease in the strength of the strong force would
lead to the nucleus being even less bound. In addition, we note that
if the strong force changes, then the binding energy of the alpha
particles will also change. In this approximation, we are assuming
that the alpha particles are much more bound than the $^8$Be
nucleus. If the strong force increases in strength, then the alpha
particles will be more tightly bound and will act as independent
entities to a greater extent. However, it is important that the
binding energy of the $^8$Be nucleus does not change at exactly the
same rate as that of two alpha particles (the previous section 
addresses this issue).  Finally, we note that the potential for the
strong force is often modeled with a Yukawa form, 
\be
V_Y = - {g^2 \over r} \exp [-\beta r] \,, 
\label{yukawa} 
\ee
where $\beta$ is an inverse length scale that sets the 
range of the strong force. In this context, we find that 
\be
{\Delta g \over g} \propto {\Delta V_0 \over V_0}\,
\ee
so that the required change in the strong force coupling strength 
is also of order a few percent.

\subsection{Bohr Model} 
\label{sec:bohr} 

Here we consider the two alpha particles that make up the $^8$Be
nucleus to be in orbit about their common center of mass and held
together by the strong force. Using the form of the Yukawa potential
from equation (\ref{yukawa}), force balance implies  
\be
m_R v^2 = {g^2 \over r} \exp[-\beta r] \left(1 + \beta r \right)\,.
\label{balance} 
\ee
The total energy has the form 
\be
E = {g^2 \over 2r} \exp[-\beta r] \left(\beta r - 1 \right)\,.
\ee
Using the Bohr quantization condition $m_R vr=n\hbar$,
the left hand side of equation (\ref{balance}) can be written 
\be
m_R v^2 = {n^2 \hbar^2 \over m_R r^2} \,.
\ee
In our universe, no bound state exists, and the energy $E$ is 
close to zero. This condition leads us to work in terms of the 
quantity 
\be
{1 \over \beta r } = 1 + \eta \,, 
\ee
where this expression serves as the definition of $\eta$. 
Since $\eta$ is small, we can replace the exponential term 
with its value in the limit $\eta\to0$ so that $\exp[\beta r]$ 
= $\econ$. The energy becomes 
\be
E = - {g^2 \beta \over 2 \econ} \eta \,,
\label{energy}
\ee
and the remaining equations can be combined to take 
the form 
\be
{\hbar^2 \beta^2 \over m_R} (1 + \eta)^2 = 
{g^2 \beta \over \econ} (2 + \eta) \,. 
\ee
Next we define the parameter 
\be
\Lambda \equiv {g^2 m_R \over \econ \beta \hbar^2} \,,  
\label{lambdadef} 
\ee
so that $\eta$ is given by the solution to 
\be
1 + 2\eta + \eta^2 = \Lambda (2 + \eta) \,. 
\label{lambdaeq} 
\ee
In our universe, the energy is essentially zero, 
so that $\eta\approx0$ and $\Lambda\approx1/2$. In general, 
the parameter $\eta$ is given by 
\be
\eta = \Lambda/2 - 1 + \left[ \Lambda^2/4 + \Lambda \right]^{1/2}\,,
\ee
where we take the positive root of the quadratic. 
Given that $\Lambda$ = 1/2 in our universe, the energy scale 
in equation (\ref{energy}) can be written 
\be
{g^2 \beta \over 2 \econ} = {\beta^2 \hbar^2 \over 4m_R} 
\approx 5.1 \, {\rm MeV} \,,
\ee
where we have used $\beta^{-1}$ = 1 fm. 

To compare with the results obtained earlier for the square well
model, let us assume that we want the bound state energy to be 1 MeV
(instead of zero). We thus need $\eta \approx 0.2$, and hence (from
equation [\ref{lambdaeq}]) $\Lambda \approx 0.65$. As a result, 
$(\Delta\Lambda)/\Lambda$ = 0.30 and $(\Delta g)/g$ = 0.15. In 
this case, a 15 percent change in the strong coupling constant 
is necessary to obtain a 1 MeV bound state. If we only require 
the bound state energy to be 0.1 MeV, then the required values 
are correspondingly smaller, $(\Delta\Lambda)/\Lambda$ = 0.03 
and $(\Delta g)/g$ = 0.015. 

\subsection{Semiempirical Mass Formula} 
\label{sec:semf} 

In this section we derive an order of magnitude estimate for the
change in the strong force necessary to make $^8$Be stable against
decay to alpha particles. This treatment uses the semiempirical mass
formula (SEMF) and is highly approximate, as the SEMF does not work
well for small nuclei (especially $^4$He).  The binding energy $E$
from the semiempirical mass formula can be written in the form 
\be
E = a_V A - a_S A^{2/3} - a_C Z^2 A^{-1/3} + a_P A^{-1/2} \,,
\ee
where $A$ is number of nucleons and where we have neglected the
asymmetry term since it vanishes for the nuclei of interest
\cite{semf,rohlf}. We want to compare the binding energy for $^8$Be 
($A$ = 8) with that for two separate $^4$He nuclei. In our universe,
$E(A=8) \approx 2E(A=4)$, although the SEMF, as written, does not
reflect this finding. Two helium nuclei have more binding energy 
than a single beryllium nucleus by a small increment $\delta$.  The
volume term (given by $a_V$) is linear in $A$ and is the same for both
states. The Coulomb term (given by $a_C$) is small and can be
neglected to the order of interest here. The condition in our universe
can thus be written 
\be
4 \left( 2^{1/3} -1 \right) a_S - \left(1 - 8^{-1/2}\right) a_P 
= -\delta \, . 
\ee
If we want to increase the difference in binding energy from 
essentially zero (as in our universe) to 1 MeV (the benchmark 
value used in previous sections), then the first term must increase
by 1 MeV. The coefficient $a_S \approx 18$ MeV (e.g., \cite{rohlf}), 
so the first term has a value of $\sim18.7$ MeV. The required change
in the coefficient is thus $(\Delta a_S)/a_S$ = 1/18.7 = 0.053. The
value of $a_S$ is determined by the strong force and is proportion to
$g^2$. As a result, the required $(\Delta g)/g \approx 0.027$. Once
again, a few percent increase in the coupling constant for the strong
force is sufficient to allow $^8$Be to be produced from alpha
particles as an exothermic reaction with yield $\sim1$ MeV.

\section{Semi-Analytic Stellar Structure Models} 
\label{sec:models} 

Stellar structure and evolution is governed by four coupled
differential equations, which describe hydrostatic equilibrium,
conservation of mass, heat transport, and energy generation
\cite{clayton,kippenhahn,hansen,chandra,phil}. These equations must 
be augmented by specification of the equation of state, the stellar
opacity, and the nuclear reaction rates. Following previous work
\cite{adams,adamsnew}, this section develops a polytropic stellar
structure model. Although approximate, the model is flexible, and
provides solutions over a range of parameter space where the constants
of nature vary by many orders of magnitude. 

It is useful to define a fundamental scale $\starmass$
for stellar masses, i.e., 
\be 
\starmass \equiv \alpha_G^{-3/2} \mpro = \left( {\hbar c \over G} 
\right)^{3/2} \mpro^{-2} \, \approx \, 
3.7 \times 10^{33} {\rm g} \approx 1.85 M_\odot \,,
\label{mscale} 
\ee 
where the numerical values correspond to standard values of the
constants. Stellar masses are comparable to this benchmark scale, in
our universe \cite{phil} and others \cite{adams,adamsnew}. The mass
scale $\starmass$ is also comparable to the Chandrasekhar mass
\cite{chandra}.  

\subsection{Hydrostatic Equilibrium Structures} 

For this model, we use a polytropic equation of state of the form
$P=K\rho^\Gamma$ where $\Gamma=1+1/n$. The polytropic index $n$ is
expected to be slowly varying over the range of stellar masses. Low
mass stars remain convective over much of their lifetimes and have
polytropic index $n$ = 3/2. On the other hand, high mass stars have
substantial radiation pressure in their interiors and the index
$n\to3$.  Use of a polytropic equation of state \cite{chandra} allows
us to replace the force balance and mass conservation equations with
the Lane-Emden equation 
\be 
{d \over d\xi} \left( \xi^2 {d f \over d \xi} \right) 
+ \xi^2 f^n = 0 \, , 
\label{laneemden} 
\ee 
where we use the standard definitions 
\be 
\xi \equiv {r \over R} , \qquad \rho = \rhocen f^n , \qquad 
{\rm and} \qquad R^2 = {K \Gamma \over (\Gamma - 1) 
4 \pi G \rhocen^{2 - \Gamma} } \, .
\label{rdef} 
\ee  
For a given index $n$, equation (\ref{laneemden}) specifies the
density profile for given values of the constants $\rhocen$ and $R$.
The corresponding pressure profile is then specified through the
polytropic equation of state.  For stars with the properties required
for nuclear fusion, the stellar material obeys the ideal gas law so
that the temperature is given by $T=P/(\rgas\rho)$, with
$\rgas=k/\mbar$. Integration of equation (\ref{laneemden}) outwards,
subject to the boundary conditions $f=1$ and $df/d\xi=0$ at $\xi$ = 0,
then determines the outer boundary of the star.  Specifically, the
stellar radius is given by $R_\ast=R\xi_\ast$, where the parameter
$\xi_\ast$ is defined to be the value of the dimensionless variable
where $f(\xi_\ast)$ = 0.

For given values of the constants $\rhocen$ and $R$, and the index
$n$, the physical structure of the star is specified.  However, for a
given stellar mass $M_\ast$, these parameters are not independent, but
rather are related through the integral constraint 
\be 
M_\ast = 4 \pi R^3 \rhocen \int_0^{\xi_\ast} \xi^2 
f^n (\xi) d\xi \, \equiv 4 \pi R^3 \rhocen \mzero \,. 
\label{mzero} 
\ee 
The second equality defines the dimensionless quantity $\mzero$,
which is a function of the polytropic index $n$ and is of order 
unity. 

\subsection{Nuclear Reactions} 

Thermonuclear fusion primarily depends on three physical variables:
the temperature $T$, the Gamow energy $E_G$, and the nuclear fusion
factor $S(E)$, where the latter quantity sets the reaction cross
section.  Here we want to determine how the nuclear ignition
temperature depends on the other variables of the problem. The Gamov
energy can be written in the form
\be
E_G = (\pi \alpha Z_1 Z_2)^2 {2 m_1 m_2 \over m_1 + m_2} c^2 \, = 
(\pi \alpha Z_1 Z_2)^2 2 m_R c^2 \, , 
\label{gamow} 
\ee 
where $m_j$ are the masses of the nuclei and $Z_j$ are the charges
\cite{clayton,hansen,phil}.  The second equality defines the reduced
mass $m_R$. For reactions involving two protons, $E_G$ = 493 keV,
whereas for two helium nuclei, $E_G$ = 31.6 MeV. The Gamow energy
specifies the degree of Coulomb barrier penetration, and is determined
by $\alpha$ (the strength of the electromagnetic force) and the masses
of the reacting particles.  The strong and weak nuclear forces
determine the reaction cross sections. In this setting, the
interaction cross sections $\sigma(E)$ are related to the nuclear
fusion factor $S(E)$ according to 
\be
\sigma(E) = { S(E) \over E} \exp \left[ - 
\left( {E_G \over E} \right)^{1/2} \right] \, , 
\label{sigma} 
\ee 
where $E$ is the energy of the interacting particles. The gas at
the center of the star obeys the ideal gas law and its constituent
particles have a distibution of energy determined by the temperature.
In ordinary stars, under most circumstances, the reaction cross
section depends on energy according to $\sigma \propto 1/E$. As a
result, the nuclear fusion factor $S(E)$ is a slowly varying function
of energy. This form for the cross section arises when the interacting
nuclei can be described by non-relativistic quantum mechanics. In this
regime, $\sigma$ is proportional to the square of the de Brogile
wavelength, so that $\sigma \sim \lambda^2 \sim (h/p)^2 \sim
h^2/(2mE)$.

The energies of the interacting nuclei have a thermal distribution, 
so that the weighted cross section has the form 
\be 
\langle \sigma v \rangle = \left[ {8 \over \pi m_R} \right]^{1/2} 
\left[ {1 \over k T} \right]^{3/2} \int_0^\infty  \sigma(E) 
\exp \left[ - E/kT \right] E dE \, . 
\label{int} 
\ee
With the thermal factor (from equation [\ref{int}]) and the coulomb
repulsion factor (from equation [\ref{sigma}]), the nuclear reaction 
rate is controlled by the composite exponential factor $\exp[-\Phi]$, 
where the function $\Phi$ includes two contributions, i.e., 
$\Phi=E/(kT)+(E_G/E)^{1/2}$.
The function $\Phi$ has a minimum value at a characteristic energy
given by $E_0 = E_G^{1/3} (kT/2)^{2/3}$.  The integral in equation
(\ref{int}) has most of its support for energies $E \approx E_0$,
where the function $\Phi$ itself takes the value 
$\Phi_0=3(E_G / 4 k T)^{1/3}$.

We can approximate the integral of equation (\ref{int}) using Laplace's 
method, so that the corresponding reaction rate $R_{12}$ for two nuclear 
can be written in the form 
\be 
R_{12} = n_1 n_2 {8 \over \sqrt{3} \pi \alpha Z_1 Z_2 m_R c (N_s!)} 
S(E_0) \Theta^2 \exp [-3 \Theta] \,,
\ee
where we have defined 
\be   
\Theta \equiv \left( {E_G \over 4 k T} \right)^{1/3}  \, . 
\label{thetadef} 
\ee
In this expression, the reacting nuclei have number densities $n_1$ and
$n_2$, and the parameter $N_S$ counts the number of identical particles
involved in the reaction. 

\subsection{Stellar Luminosity and Energy Transport} 

If we define $\varepsilon(r)$ to be the power generated per unit
volume, the luminosity of the star is determined through the equation 
\be
{d L \over d r} = 4 \pi r^2 \varepsilon(r) \,. 
\ee 
The luminosity density $\varepsilon$ can be written in terms of 
the nuclear reaction rates so that it takes the form 
\be 
\varepsilon(r) = \conlum \rho^2 \Theta^2 
\exp[-3\Theta] \, , 
\ee 
where $\Theta$ is defined in equation (\ref{thetadef}) and where 
we define a nuclear burning parameter 
\be
\conlum \equiv 
{\yield R_{12} \over \rho^2 \Theta^2} \exp[3\Theta] = 
{8\yield S(E_0) \over \sqrt{3}\pi\alpha m_1 m_2 Z_1 Z_2 m_R c (N_s!)} \,,
\label{conlumdef} 
\ee
where $\yield$ is the mean energy generated per nuclear reaction.  
For proton-proton fusion in our universe, under typical conditions,
the nuclear burning parameter $\conlum \approx 2 \times 10^4$ cm$^5$
s$^{-3}$ g$^{-1}$.  For deuterium fusion in our universe, for which
there is no bottleneck due to the weak interaction, the nuclear
constant is much larger, $\conlum \approx 2.3 \times 10^{21}$ cm$^5$
s$^{-3}$ g$^{-1}$ \cite{barnes2016}.

The total stellar luminosity $L_\ast$ is determined by integrating
over the star, 
\be
L_\ast = \conlum 4 \pi R^3 \rhocen^2 \int_0^{\xi_\ast} 
f^{2n} \xi^2 \Theta^2 \exp[-3\Theta] d\xi \, \equiv 
\conlum 4 \pi R^3 \rhocen^2 I(\thetacen) \,. 
\label{lumintegral} 
\ee 
In this expression, the second equality defines the function
$I(\thetacen)$, where we also define $\thetacen$ = $\Theta(\xi=0)$ =
$(E_G / 4 k \tcent)^{1/3}$.  Note that the temperature is given by
$T=\tcent f(\xi)$ so that $\Theta=\thetacen f^{-1/3} (\xi)$. As a
result, for a given value of the polytropic index $n$ (which
determines the form of $f(\xi)$ through the Lane-Emden equation), 
this integral is specified up to the constant $\thetacen$. 

At this stage of the derivation, the definition of equation
(\ref{rdef}), the mass integral constraint (\ref{mzero}), and the
luminosity integral (\ref{lumintegral}) provide us with three
equations for four unknowns: the radial scale $R$, the central density
$\rhocen$, the total luminosity $L_\ast$, and the coefficient $K$ in
the equation of state. The fourth equation of stellar structure is
thus required to finish the calculation. Here we are primarily
interested in larger stars where energy is transported by radiation,
so that the energy transport equation takes the form 
\be
T^3 {dT \over dr} = - {3 \rho \kappa \over 4 a c} 
{L(r) \over 4 \pi r^2} \, , 
\label{transport} 
\ee  
where $\kappa$ is the opacity (cross section per unit mass for photons
interacting with stellar material).  Next we make the following
simplification. The opacity $\kappa$ in stellar interiors generally
follows Kramer's law so that $\kappa \sim \rho T^{-7/2}$
\cite{hansen,phil}.  For polytropic equations of state, we find that
the product $\kappa \rho \sim \rho^{2 - 7/2n}$.  For the index $n=7/4$, 
the product $\kappa \rho$ is thus exactly constant. For other values
of the index $n$, the quantity $\kappa \rho$ will be slowly varying. 
As a result, we can make the approximation $\kappa\rho$ =
$\kappa_0\rhocen$ = {\sl constant} for purposes of solving the energy
transport equation (\ref{transport}). Equation (\ref{lumintegral}) 
then simplifies to the form 
\be 
L_\ast \int_0^{\xi_\ast} {\ell(\xi) \over \xi^2} d\xi = a \tcent^4 
{4 \pi c \over 3 \rhocen \kappa_0} R \, , 
\label{esolve} 
\ee
where we have defined a dimensionless luminosity density
$\ell(\xi)\equiv L(\xi)/L_\ast$ (see the integral in equation
[\ref{lumintegral}] for the full definition of $\ell(\xi)$).  For
purposes of solving equation (\ref{esolve}), we can assume that the
integrand of equation (\ref{lumintegral}) is sharply peaked toward the
center of the star. The temperature in the stellar core can be modeled
as an exponentially decaying function of position so that
$T\sim\exp[-\beta\xi]$. With these approximations (see \cite{adams} 
for further detail) the expression for $\ell(\xi)$ becomes 
\be 
\ell(\xi) = {1 \over 2} \int_0^{x_{\rm end}} x^2 {\rm e}^{-x} dx 
\, \qquad {\rm where} \qquad x_{\rm end} = \beta \thetacen \xi \, . 
\label{elldef} 
\ee 
and the luminosity (from equation [\ref{esolve}]) can be written 
in the form 
\be 
L_\ast = a \tcent^4 {4 \pi c \over 3 \rhocen \kappa_0} 
{R \over \beta \thetacen} \, . 
\label{transolve} 
\ee
Note that the parameter $\beta$ is defined implicitly.  

\subsection{Stellar Structure Solutions}  

The results of the previous section provide us with four equations and
four unknowns. These equations can be solved to obtain the following
expression for the central temperature, written here in terms of the
variable $\thetacen$,  
\be 
I(\thetacen) \thetacen^{-8} = {2^{12} \pi^5 \over 45} 
{1 \over \beta \kappa_0 \conlum E_G^3 \hbar^3 c^2} 
\left( {M_\ast \over \mzero} \right)^4 
\left[ {G \mbar \over (n + 1) } \right]^7 \,. 
\label{tcsolution} 
\ee 
Note that the right hand side of the equation is dimensionless.  The
quantities $\mzero$ and $\beta$ are dimensionless measures of the mass
and luminosity integrals over the star; these quantities, as well as
the index $n$, are of order unity. The mass of the star is $M_\ast$, 
and the remaining parameters depend on the fundamental constants. 

For typical values of the parameters in our universe, the right hand
side of this equation is approximately $10^{-9}$ for hydrogen burning
stars. For helium burning, the Gamov energy is larger by a factor of
64, and the nuclear constant is smaller, perhaps by an order of
magnitude. [Keep in mind that we are working in the counterfactual
  case where $^8$Be is stable, so that helium burning can proceed via
  two-body reactions.] The right hand side of the equation for helium
burning is thus $\sim10^{-16}$.

With the central temperature $\tcent$ (equivalently, $\thetacen$)
specified by the solution to equation (\ref{tcsolution}), we can solve
for the remaining stellar parameters. The stellar radius is given by 
\be
R_\ast = {G M_\ast \mbar \over k \tcent} {\xi_\ast \over (n+1) \mzero} \, ,
\ee 
the luminosity has the form 
\be
L_\ast = {16 \pi^4 \over 15} {1 \over \hbar^3 c^2 \beta \kappa_0 \thetacen} 
\left( {M_\ast \over \mzero} \right)^3 \left[ 
{G \mbar \over n + 1 } \right]^4 \,. 
\label{lstarsolve} 
\ee 
Finally, the photospheric temperature $T_\ast$ is determined by 
the usual outer boundary condition so that 
\be
T_\ast = \left[ {L_\ast \over 4\pi R_\ast^2\sigma_{\rm sb}} \right]^{1/4} \,, 
\label{tphoto} 
\ee 
where $\sigma_{\rm sb}$ is the Stefan-Boltzmann constant. 
 
\subsection{Minimum and Maximum Stellar Mass} 
\label{sec:minmax} 

The semi-analytic stellar structure model (developed above) can be
used to explore the range of possible stellar masses that can burn
helium in universes with stable $^8$Be. The minimum mass of a star is
determined by the onset of degeneracy pressure. For a given stellar
mass, degeneracy pressure enforces a maximum temperature that can be
attained in the stellar core. If this maximum temperature is lower
than the value required for nuclear fusion, the star cannot ignite.
Previous work \cite{adams,phil} has shown that the maximum temperature 
that can be realized in a stellar core is given by  
\be 
k T_{\rm max} = \left( 4\pi \right)^{-2/3} {5 \over 36} {m_e \over \hbar^2} 
G^2 M_\ast^{4/3} \mbar^{8/3} \, . 
\label{tmax} 
\ee 
Setting this value of the central temperature equal to the minimum
temperature $\tig$ required for nuclear burning, we obtain the minimum
stellar mass 
\be 
M_{\ast {\rm min}} = 6 (3 \pi)^{1/2} \left( {4 \over 5} \right)^{3/4}
\left( {\mpro \over \mbar} \right)^2
\left( {k \tig \over m_e c^2} \right)^{3/4} \starmass \, . 
\label{massmin} 
\ee 
Note that this minimum stellar mass is given by a dimensionless 
expression times the fundamental stellar mass scale $\starmass$ 
defined in equation (\ref{mscale}). 

By using the minimum mass from equation (\ref{massmin}) to specify the
mass in equation (\ref{tcsolution}), we can eliminate the mass
dependence and solve for the minimum value of the nuclear ignition
temperature $\tig$. The resulting temperature is given in terms of
$\thetacen$, which is given by the solution to the following equation 
\be
\thetacen I(\thetacen) = \left( {2^{23} \pi^7 3^4 \over 5^{11} } \right) 
\left( {\hbar^3 \over c^2} \right) \left( {1 \over \beta \mzero^4} \right) 
\left( {1 \over \mbar m_e^3} \right) 
\left( {G \over \kappa_0 \conlum} \right) \, . 
\label{iprofile}
\ee
The parameters on the right hand side of the equation have been
grouped to include numbers, constants that set units, dimensionless
parameters of the polytropic solution, particle masses, and parameters
that depend on the fundamental forces. The function of $\thetacen$ on
the left hand side of equation (\ref{iprofile}) is a decreasing
function of $\thetacen$ over the range of interest \cite{adams}.  
For larger values of the nuclear parameter $\conlum$, the solution to
equation (\ref{iprofile}) thus occurs at larger values of $\thetacen$
and hence lower central temperatures.  This feature implies that the
lower mass limit for helium burning can be below that for hydrogen 
burning.  The value of $\conlum$ for helium burning into stable
beryllium (Section \ref{sec:heburn}) is expected to be comparable to
that for deuterium burning in our universe, with a comparable lower
mass limit ($M_{\ast{\rm min}}\sim0.015M_\odot$ for deuterium).

For hydrogen-burning stars with zero metallicity, we can take
$\mbar=\mpro$, so that the right hand side of equation
(\ref{iprofile}) becomes {\sl\small RHS} $\approx4\times10^{-6}$,
where we have used the value of $\conlum$ appropriate for hydrogen
fusion in our universe. The function on the left-hand side of equation
(\ref{iprofile}) has a maximum value of $\sim0.05$ \cite{adams,adamsnew}. 
In order for working stellar structure solutions to exist, the
right-hand side of the equation must be less than this maximum. In our
universe, this constraint is satisfied by a factor of $\sim10^4$, and
this leeway allows (hydrogen-burning) stars to exist over a range of
possible masses ($M_\ast\sim0.1-100M_\odot$).  As outlined below, the
nuclear parameter $\conlum$ for helium burning (into stable beryllium)
is a factor of $\sim10^{16}$ larger than that for hydrogen burning in
our universe. For helium-burning stars and a pure helium composition,
we can take $\mbar=4\mpro$ and the {\sl\small RHS} $\sim 10^{-22}$.
Because of the larger value of the nuclear parameter $\conlum$, stars
burning helium into beryllium can exist over a wider range of 
parameter space than those burning hydrogen. Note that this result is
not unexpected --- in our universe, the stellar mass range for burning
deuterium (which has a larger value of $\conlum$) is wider than that
for hydrogen.  In general, this constraint can be written in the form 
\be
\left( {G\over G_0} \right)  
\left( {\alpha\over\alpha_0} \right)^{-2} 
\left( {\conlum\over \conlum_0} \right)^{-1} < 1.2 \times 10^4 \,, 
\ee
where $\conlum_0 = 2 \times 10^4$ (in cgs units). 

An estimate for the maximum stellar mass can also be obtained. In
massive stars, the central pressure has contribution from gas pressure
(through the ideal gas law) and from radiation pressure. If the
radiation pressure dominates over the gas pressure, then the star does
not have a stable hydrostatic state available to it. In the limit
where the radiation pressure is large, the star has the same energy
(self-gravity and thermal) for any radial size and thus cannot be
stable \cite{phil}. If we let $\bcon$ be the fraction of the central
pressure provided by gas, the maximum stellar mass has the form 
\be
M_{\ast {\rm max}} = \left( {36 \over \pi} \right)^{1/2}
\left[ {3 \over a} {(1 - \bcon) \over \bcon^4} \right]^{1/2} 
G^{-3/2} \left( {k \over \mbar} \right)^2 \, =  
\left( {36 \sqrt{10} \over \pi^{3/2} } \right) 
\left( {\mpro \over \mbar} \right)^2 \, M_0 \,, 
\label{maxmass}
\ee
where the second equality assumes that $\bcon=1/2$. This value is
often used, although the stability boundary is not a perfectly sharp
function of $\bcon$. With this choice, the above expression becomes
$M_{\ast {\rm max}} \approx 20 (\mpro/\mbar)^2 M_0$.  Since massive
stars are highly ionized, $\mbar \approx 0.6 \mpro$ under standard
conditions, and hence $M_{\ast {\rm max}} \approx 56 M_0 \approx 100
M_\odot$ for our universe. Note that this limit is nearly identical to
the constraint derived from the Eddington luminosity \cite{adams}.
Notice also that this upper limit on stellar mass is independent of
the nuclear burning parameter $\conlum$ (to leading order), so that
alternate universes are expected to have the same cutoff (for fixed
values of $G$ and $\mpro$).

\subsection{Helium Burning into Beryllium} 
\label{sec:heburn} 

We now consider stellar structure models for universes where beryllium
has a stable $A$ = 8 isotope. In this scenario, after hydrogen burning
has run its course, the helium that has built up in the stellar core
can burn directly into $^8$Be. We thus assume that the reaction 
\be 
^4{\rm He} + ~^4{\rm He} \to ~^8{\rm Be} + \gamma 
\label{heburn} 
\ee
is viable. In order to explore the properties of stars fueled by this
reaction, we use the stellar model developed in previous sections. 

As a start, we set the structure parameters $\alpha$ and $\alpha_G$ to
have the standard values in our universe. For varying choices of the
nuclear parameters, we can then compare the properties of these helium
burning stars to those in our universe. The composition of the star
must also be specified. Here we consider the stellar core to be
composed entirely of helium, i.e., we assume that hydrogen burning in
the core has proceeded to completion. 

Because the reaction (\ref{heburn}) does not occur in an exothermic
manner in our universe, the reaction rate cannot be measured, but we
still need to specify the nuclear parameter $\conlum$.  Keep in mind
that this quantity includes both the energy per reaction $\Delta E$
and the nuclear fusion factor $S(E_0)$ (see equation
[\ref{conlumdef}]). Since the reaction (\ref{heburn}) takes places
through the strong force, we expect the reaction rate to be relatively
large. More specifically, the proton-proton chain for hydrogen burning
(in our universe) requires the weak interaction, as two protons must
be converted into neutrons to make helium, and $\conlum \sim 10^4$ in
cgs units.  For comparison, for deuterium burning reactions the weak
force does not come into play, and $\conlum \sim 10^{22}$ in the same
units.  Although the energy per reaction $\Delta E$ for helium burning
(here into beryllium) is likely to be smaller than that for helium
production in our univese, the difference is likely to be less than a
factor of ten; the lack of a weak interaction bottleneck should more
than compensate.  We thus expect the nuclear parameter $\conlum$ for
$2\alpha\to$~$^8$Be to be closer to that for deuterium burning
($\conlum \sim 10^{22}$ cgs) than for hydrogen burning ($\conlum \sim
10^{4}$ cgs).  Given the uncertainty in this parameter, we explore a
range of values, from $\conlum = 10^{12}$ to $10^{24}$ cm$^5$ sec$^{-3}$
g$^{-1}$.

\begin{figure}[tbp]
\centering 
\includegraphics[width=.90\textwidth,trim=0 150 0 150,clip]{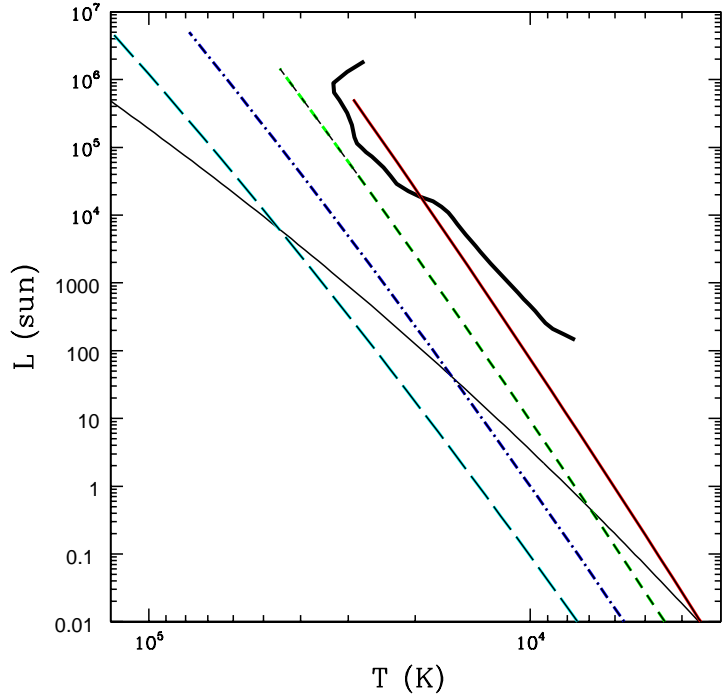}
\caption{H-R digram for stars that can burn helium into beryllium. 
The colored curves show the main-sequence for helium burning in
alternate universes where $^8$Be is stable, for four choices of the
nuclear burning parameter, from left to right: $\conlum$ (in cgs
units) = $10^{12}$ (cyan), $10^{16}$ (blue), $10^{20}$ (green), and
$10^{24}$ (red). For comparison, the thick black curve shows the
helium burning main-sequence for stars in our universe where the
triple alpha process is active. The thin black curve shows the main
sequence for hydrogen burning stars with a single $p$-$p$ nuclear
reaction chain and the structure of an $n$ = 3/2 polytrope. }
\label{fig:hrdiagram} 
\end{figure}  

The H-R diagram for these helium burning stars is shown in Figure
\ref{fig:hrdiagram}, where these results are obtained using the
semi-analytic stellar structure model. The locations of the helium
burning main-sequence are shown for four choices of the nuclear
parameter $\conlum$ (where the spacing of the values is even in the
logarithm). For each value of $\conlum$, the range of allowed stellar
masses is different (see Section \ref{sec:minmax}).  Note that these
helium burning stars occupy approximately the same region of the H-R
diagram as stars in our univese. In Figure \ref{fig:hrdiagram}, the
thin black curve shows the hydrogen burning main-sequence for stars
operating with only a single proton-proton nuclear reaction chain
(calculated using the semi-analytic model). As another comparison, the
thick black curve shows the helium burning main-sequence for stars in
our universe (calculated numerically as described in Section
\ref{sec:mesa}). This curve varies with time and corresponds to the
condition that the carbon abundance in the stellar core has reached
2\%.  In our universe where the triple alpha process operates, helium
burning is suppressed by the need for three-body reactions, but is
enhanced due to the Hoyle resonance. Compared to the alternate
scenario explored here, helium burning stars in our univese have
somewhat higher central temperatures and luminosities. Nonetheless,
the luminosities are roughly comparable to those of stars in our
universe.

\begin{figure}[tbp]
\centering 
\includegraphics[width=.90\textwidth,trim=0 150 0 150,clip]{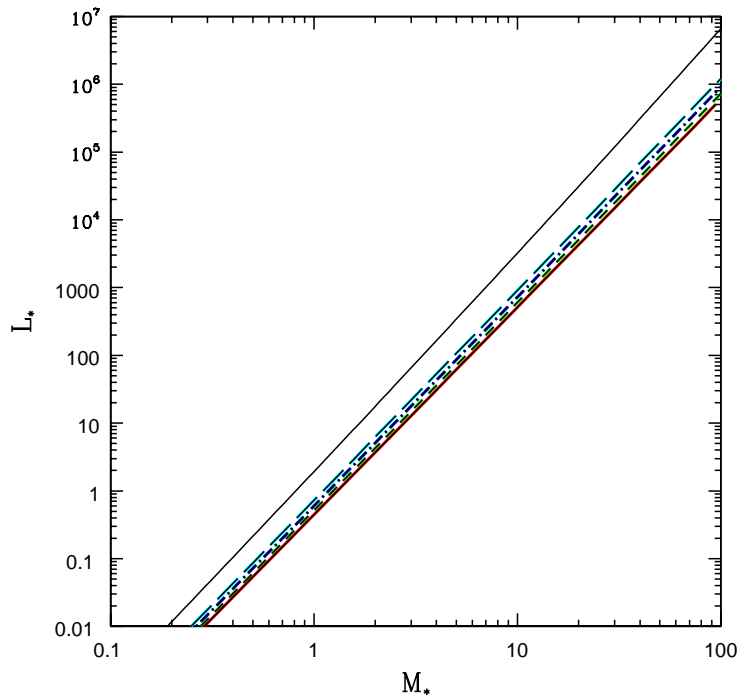}
\caption{Luminosity versus stellar mass relation for stars that burn 
helium into beryllium. The curves show the luminosities for helium
burning stars in alternate universes where $^8$Be is stable, for four
choices of the nuclear burning parameter, from top to bottom: 
$\conlum$ (in cgs units) = $10^{12}$ (cyan), $10^{16}$ (blue),
$10^{20}$ (green), and $10^{24}$ (red). For comparison, the upper
black line shows the relation for hydrogen burning stars using the
simplified stellar model (see text). }
\label{fig:masslum} 
\end{figure} 

The stellar luminosity is plotted versus stellar mass in Figure
\ref{fig:masslum}.  Note that the allowed mass range for helium
burning stars extends down to much lower masses than for the hydrogen
burning stars in our universe. This extension is a direct result of
the larger value of $\conlum$. With higher $\conlum$, nuclear
reactions can take place at lower temperatures, which can be achieved
with smaller stellar masses. This effect is modest --- the variation
in stellar mass range is much smaller than the variation in $\conlum$
--- because of the extreme sensitivity of nuclear reactions to
temperature. Deuterium burning in our universe also proceeds via the
strong force, with a large value of $\conlum$, and deuterium burning
extends down to $M_\ast \approx 0.015 M_\odot$ (small compared to the
minimum mass of about 0.1 $M_\odot$ for hydrogen burning). Except for
the differences in the allowed mass range, the mass-luminosity
relation is similar for all values of $\conlum$ and for hydrogen
burning stars.  The variation in the luminosity with $\conlum$ for a
given mass is much smaller than the variation in luminosity with mass.

\begin{figure}[tbp]
\centering 
\includegraphics[width=.90\textwidth,trim=0 150 0 150,clip]{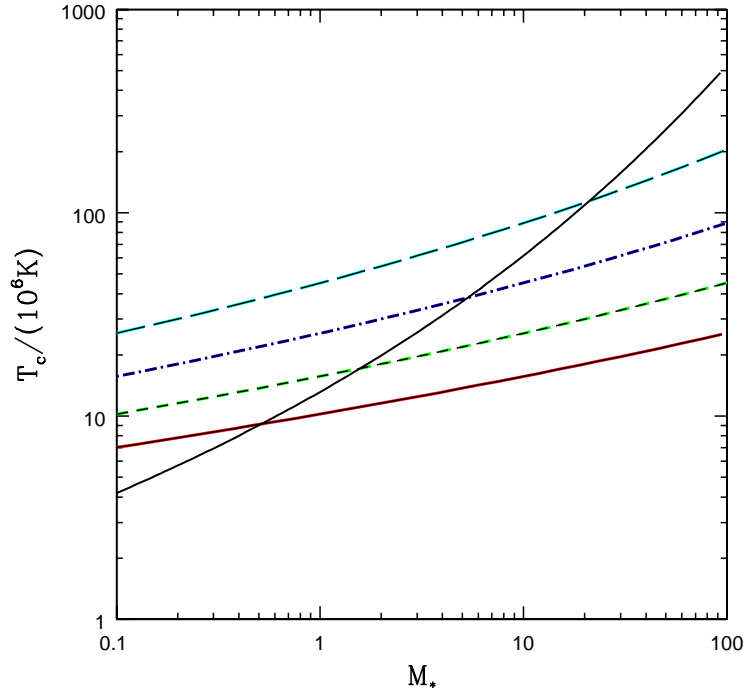}
\caption{Central temperature versus stellar mass relation for stars 
that burn helium into beryllium. The colored curves show the temperatures 
for helium burning stars in alternate universes where $^8$Be is
stable, for four choices of the nuclear burning parameter, from top to
bottom: $\conlum$ (in cgs units) = $10^{12}$ (cyan), $10^{16}$ (blue),
$10^{20}$ (green), and $10^{24}$ (red). For comparison, the black curves
shows the central temperature as a function of mass for stars burning 
hydrogen (through the p-p chain). }
\label{fig:centemp} 
\end{figure} 

For completeness, the central temperatures of the helium burning stars
are shown in Figure \ref{fig:centemp}. Again, the range of stellar
masses extends to lower values, but the range of central temperatures
is roughly comparable to those of hydrogen burning stars in our
universe, from a few million to a few hundred million Kelvin. The
slope of the central temperature curve is less steep for helium
burning stars. This difference arises due to the greater temperature
sensitivity --- a small change in central temperature $T_{\rm c}$
leads to an enormous change in the nuclear reaction rates. As a
result, the range of central temperatures for a given value of the
nuclear parameter $\conlum$ is smaller for helium burning stars
(compared to hydrogen burning stars in our universe).

Taken together, Figures \ref{fig:hrdiagram} -- \ref{fig:centemp} show
that stars burning helium into stable beryllium have properties that
are roughly comparable to ordinary hydrogen burning stars in our
universe. 

\section{Numerical Stellar Evolution Models} 
\label{sec:mesa} 

The previous section demonstrates that stars in other universes can
successfully burn helium into beryllium, provided that that latter
nucleus has a stable state. These results were obtained using an
extremely simple stellar model in order to understand the process at
the most basic level. Since the regime of operation for these stars is
quite different from those of ordinary stars in our universe, this
section pursues the issue further by considering stellar evolution
models using the state-of-the-art computational package 
{\sl\small MESA} \cite{paxton,paxtonb}. The semi-analytic model is
robust because of its simplicity. However, the model uses only a
single nuclear reaction (at a given time), considers only stellar
structure rather than time evolution, and is hard-wired with a
polytropic equation of state. In contrast, the {\sl\small MESA}
package can evolve the stars from before their pre-main-sequence
phase, onto the hydrogen burning main-sequence, through helium
burning, and beyond. The code includes contributions to the equation
of state from degeneracy pressure, radiation pressure, and gas
pressure; it also includes both convective and radiative energy
transport.

\subsection{Modifications to the {\sl\small MESA} Numerical Package}
\label{sec:mesachanges} 

In order to use {\sl\small MESA} to study $^8$Be production in other
universes, we had to modify two principal components of the code.
First, we needed to include a stable $^8$Be isotope. In our universe,
the binding energy of $^8$Be, denoted here as $B_8$, is given by 
\be
  B_8 =[4m_p + 4m_n - M(^8{\rm Be})]c^2,
\ee
where $M$($^8$Be) the mass of $^8$Be, $m_p$ is the mass of a proton,
and $m_n$ is the mass of a neutron.  With the standard values, the
binding energy is $B_8\simeq56.5\,{\rm MeV}$ \cite{audi}.  For
comparison, the binding energy of $^4$He is 
\be
  B_4 =[2m_p + 2m_n - M(^4{\rm He})]c^2\simeq28.296\,{\rm MeV}.
\ee
As a result, the binding energy of two $^4$He nuclei is larger than that
of a single $^8$Be nucleus by $\sim100\,{\rm keV}$.  Within the
isotope list of {\sl\small MESA}, we changed the mass of $^8$Be so that 
its binding energy increses to $B_8\simeq56.8$ MeV.  We preserved the
mass of $^4$He so that a single $^8$Be nucleus has a larger binding
energy than two $^4$He nuclei by $\sim200\,{\rm keV}$.  With the 
change in binding energies, $^8$Be no longer decays into two $^4$He
nuclei and therefore subsists in appreciable amounts in stellar
interiors.

The second component that we modified in {\sl\small MESA} is the
averaged product of cross section and speed, denoted
$\langle\sigma{v}\rangle$, for the new reactions. Because $^8$Be is
unstable in our universe and decays on a time scale much shorter than
stellar evolution time scales, previous treatments included neither
the nuclide nor any reactions associated with that nuclide. Instead, 
a small equilibrium abundance of unstable $^8$Be is built up in the 
core, and some fraction immediately is converted into $^{12}$C, so 
that the nuclear properties of $^8$Be effectively cancel out (for 
further detail, see \cite{clayton,hansen,kippenhahn}). In the 
alternate universes under consideration here, the isotope $^8$Be 
is stable, so that we no longer need the triple alpha ($3\alpha$)  
channel to synthesize $^{12}$C from $^4$He. As a result, we broke
apart the $3\alpha$ reaction into two separate reactions, where the
first one synthesizes beryllium, 
\be
^4{\rm He} + \, ^4{\rm He} \quad \leftrightarrow \quad
^8{\rm Be} + \gamma,
\label{eq:r_4_4_to_8} 
\ee
and the second reaction converts beryllium into carbon, 
\be
^4{\rm He} + \, ^8{\rm Be} \quad \leftrightarrow \quad
^{12}{\rm C} + \gamma.
\label{hetobe} 
\ee
Unlike the case of the triple alpha reaction in our universe, the 
second reaction (\ref{hetobe}) can take place at any later time, 
and even in a different star (in a different stellar generation). 
We thus turn our attention to the reaction in equation
(\ref{eq:r_4_4_to_8}), which we denote as $\aabe$ $^8$Be.  We provide
the motivation for our choice of $\langle\sigma{v}\rangle$ for $\aabe$
$^8$Be by considering the $3\alpha$ reaction.  The cross section
$\langle\sigma{v}\rangle$ for $3\alpha$ is the sum of two constituent
parts, a resonant component and and a nonresonant component. The
resonance channel is not relevant for our studies, so we focus on the
nonresonant channel.  Reference\ \cite{nomoto} calculates an analytic
approximation for $\langle\sigma{v}\rangle$ in the nonresonant 
channel of $\aabe$ $^8$Be, denoted here as 
$\langle\alpha\alpha\rangle^{\star}$, with the following form: 
\be
\langle\alpha\alpha\rangle^{\star}= 6.914\times10^{-15}\,{\rm cm^3/s}\,
\,\,T_9^{-2/3} \exp(-13.489T_9^{-1/3}) \qquad \qquad \qquad 
\label{eq:aa}
\ee
$$
\qquad 
\times\,(1 + 0.031T_9^{1/3} + 8.009T_9^{2/3} + 1.732T_9
+ 49.883T_9^{4/3} + 27.426T_9^{5/3})
$$
where $T_9$ is the temperature in units of $10^9{\rm K}$.  Using
equations (\ref{int}) and (\ref{conlumdef}), we can calculate the
$S(E_0)$ factor from equation (\ref{eq:aa}), which leads to the result 
\be
S(E_0) \approx 6.760\times10^{-31} {\rm erg}\,{\rm cm}^{2} 
\approx 422\,{\rm keV}\,{\rm barn}\,.
\ee
Equation\ (\ref{conlumdef}) defines the nuclear burning parameter 
$\conlum$, which includes both the nuclear fusion factor $S(E_0)$ and 
the mean energy per reaction $\Delta E$. For the reaction $\aabe$ $^8$Be, 
the nuclear burning parameter takes the following form: 
\be\label{eq:caa}
  \conlum^{\star} = \frac{8\langle\Delta E\rangle^{\star}
  S(E_0)}{\sqrt{3}\pi\alpha m_\alpha^2Z_\alpha^2m_Rc}\,,
\ee
where $\langle\Delta E\rangle^{\star}=2B_4-B_8$ is the endothermic
energy ``yield'' for 2($^4$He) $\rightarrow$ $^8$Be.  After
substituting the values from our universe into equation (\ref{eq:caa}), 
we find $\conlum^{\star}=5.69\times10^{23}{\rm cm}^5$ s$^{-3}$
g$^{-1}$.  When calculating the cross section for the triple alpha
process in a reaction network, it is common to multiply
equation (\ref{eq:aa}) by additional factors in order to fit to
experimental data (see Ref.\ \cite{caughlan85,caughlan88}). For our 
purposes, we do not include the fitting factors, but instead include
an overall constant.  As a result, our expression for
$\langle\sigma{v}\rangle$ in alternate universes can be written in the
form 
\be
  \langle\alpha\alpha\rangle=\frac{{\conlum}}{5.69\times10^{23}}
  \left(\frac{\langle\Delta E\rangle^\star}{\langle\Delta E\rangle}\right) 
  \langle\alpha\alpha\rangle^\star \,,
\label{eq:sigvbarC}
\ee 
where $\Delta E$ (without the superscript) is the positive energy
difference $\Delta E = B_8 - 2B_4$ for the universe in question (and
where the expression is in cgs units). We incorporated equation\ 
(\ref{eq:sigvbarC}) into the {\sl\small MESA} reaction library and
added the reaction (\ref{eq:r_4_4_to_8}) to the network.  Note that
this procedure does not preserve unitarity for the $^8$Be compound
nucleus.  To preserve unitarity, the differential cross sections for
all of the reactions that include a compound $^8$Be nucleus would also
have to be modified.  Such a task is difficult and is beyond the scope
of this present work (see Ref.\ \cite{paris} for further details about
unitarity in nuclear reaction networks).

\subsection{Results for Helium Burning into Beryllium} 
\label{sec:mesaresults} 

This section presents the results obtained for a range of possible
nuclear reaction rates for $^8$Be production.  As outlined above, in
order to run a version of {\sl\small MESA} with the new $^8$Be
physics, we added $^8$Be to the list of isotopes and $\aabe$ $^8$Be to
the reaction list. Specifically, the list of isotopes includes the
following: $^1{\rm H}$, $^3{\rm He}$, $^4{\rm He}$, $^8$Be, $^{12}$C,
$^{14}{\rm N}$, $^{16}{\rm O}$, $^{20}{\rm Ne}$, and $^{24}{\rm Mg}$.
The resulting reaction list contains 18 reactions, including the
reaction for $\aabe$ $^8$Be but excluding the $3\alpha$ reaction. 
With this restricted list of isotopes and reactions, we can study 
helium burning into beryllium, as well as the production of carbon 
and oxygen. Note that the inverse reactions must also be included. 
The production of heavier elements (e.g., iron) is not considered 
here. 

These simulations were performed using the initial value $Z=10^{-4}$
for the metallicity of the star. This relatively low value of
metallicity allows us to more easily interpret the production of heavy
nuclei through stellar processes and to compare results with the
simple model of the previous section. On the other hand, for technical
reasons, the numerical code runs more robustly for $Z\ne0$.  The
elemental abundances for the more common isotopes for this choice of
metallicity ($Z=10^{-4}$) are listed in Table \ref{tab:massfract}. 
To obtain these abundances, we started with the values given in 
Ref.\ \cite{lodders} for the metals that are stable in our universe
(those nuclei with atomic number $>2$ excluding $^8$Be) and scaled
them such that they preserved their relative abundances and summed up
to $95\%$.  We then added in an ad-hoc $^8$Be mass fraction of $5\%$.
Finally, we multiplied all of the metal mass fractions by $10^{-4}$
and added them to the hydrogen and helium mass fractions to obtain the
resulting set of abundances in Table \ref{tab:massfract}.

The other two parameters that must be specified are the stellar mass
$M_\ast$ and the nuclear burning parameter $\conlum$ for the
production of $^8$Be. For the sake of definiteness, we consider the
range of stellar masses to be similar to that of our universe, i.e.,
$M_\ast\sim0.1-100 M_\odot$. Due to convergence issues, we sometimes
use a somewhat smaller range of masses. Because the cross section for
helium burning ($^8$Be production) is unknown, we consider a wide
range for the nuclear burning parameter $\conlum=10^{16}-10^{24}$ (in
cgs units). The upper end of this range is comparable to that
appropriate for deuterium burning in our universe, which proceeds
rapidly because the reaction only involves the strong force. For
comparison, hydrogen burning in our universe is characterized by a
much smaller effective value of $\conlum$ (specifically, $\conlum
\sim10^4$ in cgs units for the p-p reaction chain). The smaller value
for hydrogen burning reflects the fact that the nuclear reaction chain
must convert two protons into neutrons (thereby involving the weak
force) in order to synthesize helium.

\begin{table*}
\begin{center} 
  \begin{tabular}{cc}
    \hline
    \hline
    Isotope & Mass Fraction \\
    \hline
    $^1{\rm H}$ & 0.76 \\ 
    $^3{\rm He}$ & $3.0\times10^{-5}$ \\ 
    $^4{\rm He}$ & 0.24 \\ 
    $^8{\rm Be}$ & $4.9\times10^{-6}$ \\ 
    $^{12}{\rm C}$ & $1.7\times10^{-5}$ \\ 
    $^{14}{\rm N}$ & $4.8\times10^{-6}$ \\ 
    $^{16}{\rm O}$ & $4.5\times10^{-5}$ \\ 
    $^{20}{\rm Ne}$ & $1.0\times10^{-5}$ \\
    $^{24}{\rm Mg}$ & $2.0\times10^{-5}$ \\ 
    \hline
    \hline
  \end{tabular}
\end{center}  
  \caption
  {\label{tab:massfract} 
Table of initial mass fractions for nonzero metallicity. (Note that
the values for H and $^4$He are slightly smaller than those listed so
that the total adds up to unity.) }
\end{table*}

For given choices of metallicity, stellar mass, and nuclear burning
parameter, the {\sl\small MESA} code evolves the star from an initial
state, down its pre-main-sequence track, onto the main-sequence where
it burns hydrogen into helium, and then through the helium burning
phase.\footnote{Note that stars take $\sim0.1-0.2$ Myr to form
  \cite{fatuzzo}, at least in our universe, so that the first stages
  of this evolutionary sequence are not physically realistic. In
  particular, the pre-main-sequence evolutionary time scale is shorter
  than the formation time for stars more massive than $M_\ast \sim 7
  M_\odot$, so that massive stars do not have a pre-main-sequence
  phase.}  With the reactions considered here, including a stable
$^8$Be isotope, the stars reach a helium burning main-sequence
analogous to that found in the previous sections using the
semi-analytic model. 

\begin{figure}
\begin{center}
\includegraphics[scale=0.60]{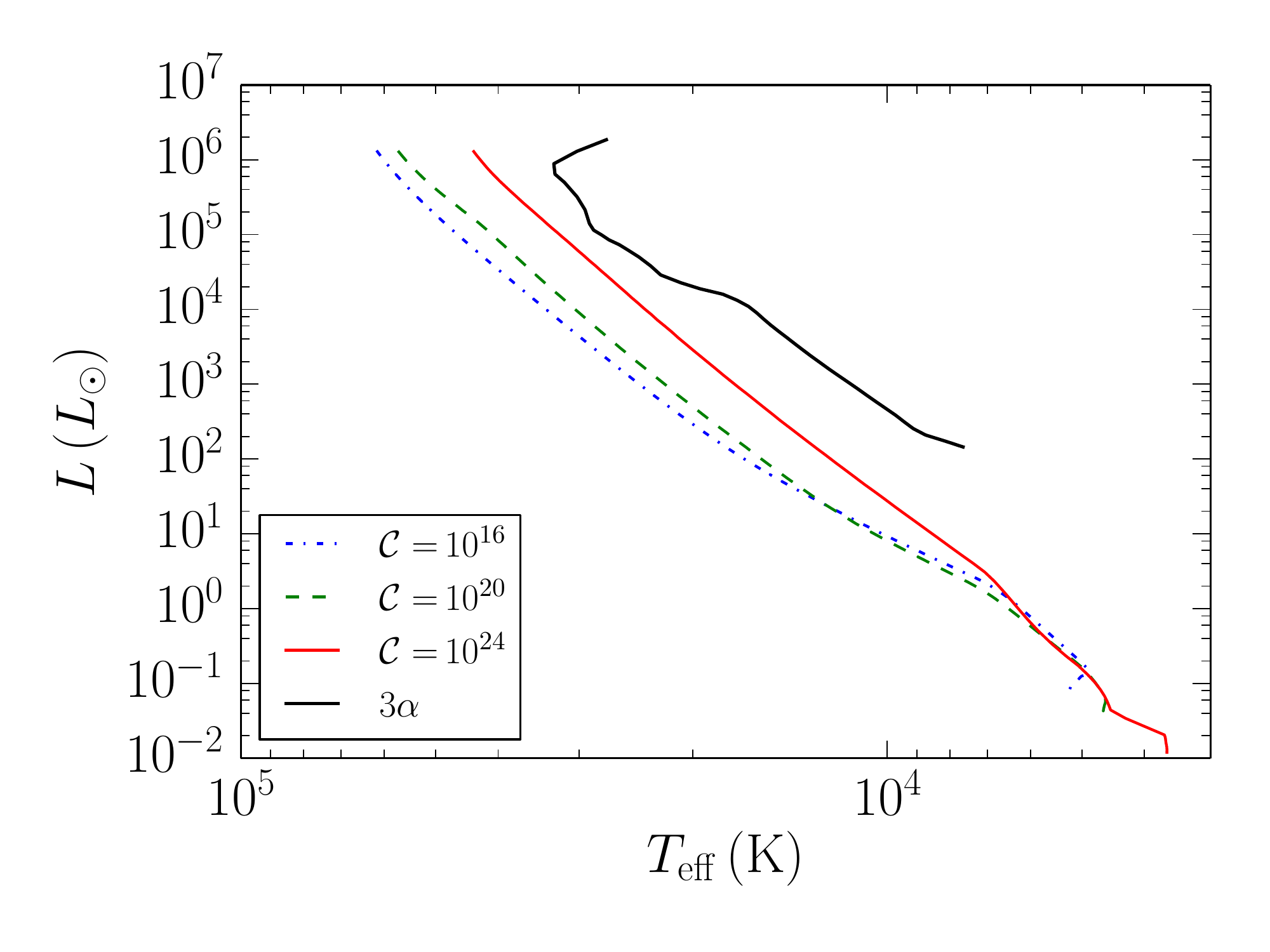}
\end{center}
\caption{H-R diagram at the start of $^4$He burning for universes  
with different values of $\conlum$ (values given in cgs units), from
left to right: $10^{16}$ (blue), $10^{20}$ (green), and $10^{24}$
(red).  Initially, the stars have metallicity $Z=10^{-4}$. In these
simulations, the start of helium burning is defined to occur when the
amount of $^8$Be rises to 2\% in the core. For comparison, the solid 
black curve shows the main-sequence for $^4$He burning via the triple
alpha process. The start of helium burning for the triple alpha curve
occurs when the $^{12}$C abundance reaches 2\% in the stellar core. }
\label{fig:hr-mesa} 
\end{figure}

Figure \ref{fig:hr-mesa} shows resulting Hertzsprung-Russell (H-R)
diagram for a range of values for the nuclear parameter $\conlum$. In
the figure, the curves represent the zero-age-main-sequence for 
helium burning (into beryllium). The onset of helium burning is not
perfectly sharp in time, so that we need to specify the criterion used
to define the helium burning main-sequence. The main-sequences shown
in Figure \ref{fig:hr-mesa} correspond to the epoch when the abundance
of $^8$Be reaches $2\%$ in the stellar core, where the core is defined
to be the inner $10\%$ of the star by mass.  Note that these
main-sequences are in good qualitative agreement with those calculated
from the semi-analytic model, as shown in Figure \ref{fig:hrdiagram}.
Notice also that all stars must get brighter as they burn through
their nuclear fuel supply, and that the helium burning phase is much
shorter than the hydrogen burning phase. As a result, the location of
the main-sequence varies with time, much more than the case of
hydrogen burning.

The H-R diagram of Figure \ref{fig:hr-mesa} also includes the main
sequence for helium burning in our universe using the standard
$3\alpha$ process (for metallicity $Z=10^{-4}$).  For this $3\alpha$
main-sequence, shown as the solid black curve, the results are plotted 
at the epoch when the mass fraction of $^{12}$C in the core reached 
$2\%$. Note that for this regime of low metallicity, the $3\alpha$
reaction has a nuclear burning parameter with an effective value
$\conlum^\star$ $\approx6\times10^{23}$ in cgs units (from equation
[\ref{eq:caa}]).  As a result, the helium burning main-sequence
occupies roughly the same region of the H-R diagram for the $3\alpha$
process in our universe and the beryllium producing process in other
universes. In detail, however, stars burning helium in our universe
have somewhat lower surface temperatures than the $\conlum\sim10^{24}$
curve for other universes, i.e., the helium burning main-sequence in
our universe falls to the right of those in alternative universes that
produce $^8$Be. One reason for this difference is the composition of
the stellar core. In our universe, the stars burn most of the hydrogen
into helium before the onset of the triple alpha process. For large
values of $\conlum$ in other universes, however, helium burning starts
while the core maintains a significant fraction of hydrogen. 

The location of the main-sequence found here for the $3\alpha$ process
depends sensitively on the stage of evolution when the
``main-sequence'' is defined.  The tracks of massive stars move back
and forth as they evolve beyond the (hydrogen-burning) main-sequence.
As a result, for a fixed stopping condition in the numerical code, the
resulting main-sequence for the $3\alpha$ process suffered from minor
numerical artifacts at both low and high stellar masses. The tracks in
the H-R diagram displayed complicated non-monotonic behavior, with
small oscillations superimposed on the otherwise smooth curve. To
address this issue, we employed a smoothing algorithm to obtain the
black solid curve shown in Figure \ref{fig:hr-mesa}. The raw data
contained points that fell to the right side (lower temperatures) of
the black curve.  The smoothing algorithm removed those points,
essentially yielding an envelope, which thus provides an upper limit
to the temperature for the helium burning main-sequence. Notice also
that helium burning in our universe occurs much more rapidly than
hydrogen burning, so that stars move relatively quickly in the H-R
diagram. The location of the helium burning main-sequence thus depends
on its definition (taken here to be when the core has reached a
composition of 2\% carbon).

\begin{figure}
\begin{center}
\includegraphics[scale=0.60]{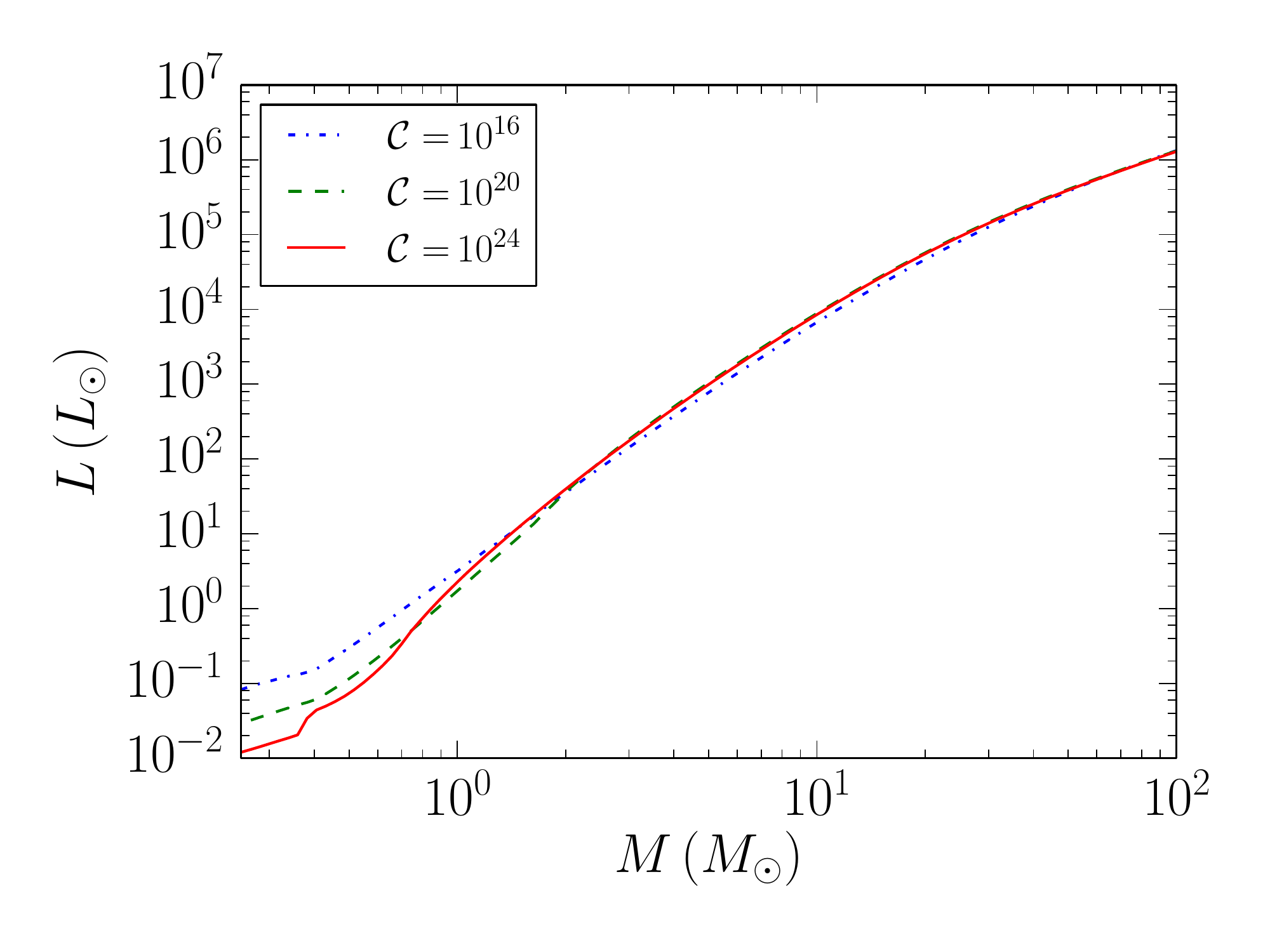}
\end{center}
\caption{Luminosity-mass relations at the start of $^4$He burning for  
universes with different values of $\conlum$ (in cgs units), from top
to bottom: $10^{16}$ (blue), $10^{20}$ (green), and $10^{24}$ (red).
Initially, the stars have metallicity $Z=10^{-4}$.}
\label{fig:lm-mesa}  
\end{figure}

Figure \ref{fig:lm-mesa} shows the mass versus luminosity
relationships for helium burning stars in other universes, again for
starting metallicity $Z=10^{-4}$. For these stars, the luminosity
at a given mass does not vary appreciably with the nuclear burning
parameter $\conlum$, which is varied over 8 orders of magnitude in the
figure. For any value of $\conlum$, the luminosity of the star must
adjust so that its energy generation ultimately supplies enough
pressure to hold up the star against its self-gravity. The stellar
mass and the opacity (which determines the energy loss rate) is the
same for all values of $\conlum$, so the luminosity is slowly varying
(as a function of $\conlum$).  Moreover, to leading order, all of the
curves show the expected power-law scaling $L_\ast\propto M_\ast^3$,
which is essentially the same as that found using the semi-analytic
model (see Figure \ref{fig:masslum}). A more detailed comparison shows
that the full stellar evolution code (the {\sl\small MESA} results
shown in Figure \ref{fig:lm-mesa}) produces a mass-luminosity relation
with more curvature than that of the semi-analytic model.  This
curvature results from the more complicated physics included in the
numerical code (see below).

\begin{figure}
\begin{center}
\includegraphics[scale=0.60]{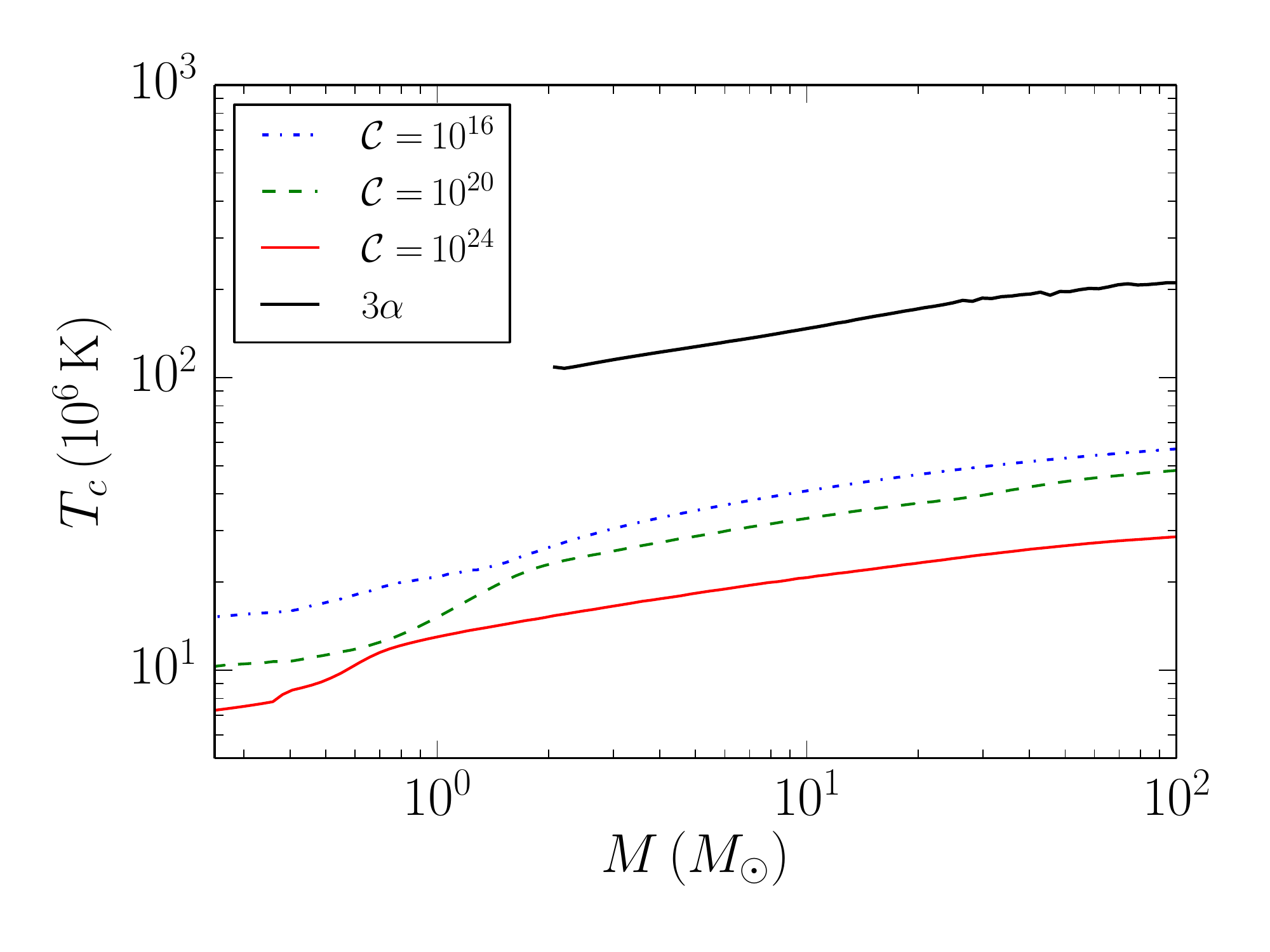} 
\end{center}
\caption{Central-temperature plotted as a function of stellar mass 
at the start of $^4$He burning for universes with different values of
$\conlum$ (in cgs units), from top to bottom: $10^{16}$ (blue),
$10^{20}$ (green), and $10^{24}$ (red). Initially, the stars have
metallicity $Z=10^{-4}$. For comparison, the central temperature for 
stars in our universe, operating via the triple alpha reaction, is 
shown as the solid black curve. }
\label{fig:tcm-mesa} 
\end{figure}

Figure \ref{fig:tcm-mesa} shows the central temperatures for helium
burning stars as a function of stellar mass, for initial metallicity
$Z=10^{-4}$.  For these stars, the central temperatures vary from
about $10^7$ K to just under $10^8$ K accross the range of stellar
masses. Stars with lower values of the nuclear burning parameter
$\conlum$ require higher central temperatures to produce (almost) the
same luminosity (see Figure \ref{fig:tcm-mesa}), i.e., the higher
central temperature compensates for the lower reaction cross section.
As expected, larger stars require higher central temperatures to
support their higher mass. These same general trends are indicated by
the semi-analytic stellar model (compare with Figure
\ref{fig:centemp}). Figure \ref{fig:tcm-mesa} also shows the central
temperature during helium burning for stars in our universe operating
via the triple alpha process (black solid curve). Note that the
central temperatures for these stars are higher by factors of
$\sim4-10$ compared to stars that produce stable $^8$Be.  In our
universe, stars thus have to reach higher central temperatures for the
triple alpha process to operate efficiently. The black curve does not
extend down to the lowest masses. Stars in this regime experience a
helium flash \cite{kippenhahn}, and this complication makes it
difficult to define the helium burning main-sequence in the same
manner as for larger stars; nonetheless, helium burning in these 
stars takes place at a central temperature $T_c\sim10^8$ K 
\cite{clayton,kippenhahn}, which corresponds to an extension of 
the curve. 

Finally, we note that the semi-analytic model and the detailed
numerical package of {\sl\small MESA} produce results are that are in
excellent qualitative and good quantitative agreement.  For a wide
range of values for the nuclear parameter $\conlum$, which includes
the cross sections and yields for helium burning into $^8$Be, both
approaches produce helium burning main-sequences that occupy the same
region of the H-R diagram (compare Figure \ref{fig:hrdiagram} with
Figure \ref{fig:hr-mesa}).  Both approaches also predict mass versus
luminosity relations of the basic form $L_\ast\propto M_\ast^3$
(compare Figure \ref{fig:masslum} with Figure
\ref{fig:lm-mesa}). Finally, the central temperatures of the stars, as
a function of stellar mass, are also similar for the two approaches
(compare Figure \ref{fig:centemp} with Figure \ref{fig:tcm-mesa}). The
differences arise due to the more complicated physics that is included
in the numerical model. Because the semi-analytic model uses a simple
polytopic equation of state, the physical structure of the star is
decoupled from its thermodynamic processes and the resulting stars
have limited forms available. In contrast, the numerical model has
many more degrees of freedom. For example, during the evolution of
high mass stars ($M_\ast=30-100 M_\odot$), the equation of state
contains contributions from both radiation pressure and ordinary gas
pressure. Since stars dominated by radiation pressure have the same
energy with different radial sizes \cite{hansen,phil}, such stars are
close to instability. They pulsate as they evolve, and thus cycle
through different radii; as a result, their surface temperatures vary
significantly with time and with stellar mass. In addition, such stars
are often in a state where the energy transport mechanism alternates 
between being convective and radiative. This variation, which is
included automatically in the numerical treatment of {\sl\small MESA},
would correspond to different choices for the polytropic index $n$ in
the semi-analytic model, but $n$ is not allowed to vary. These
complications, and others, thus lead to the modest differences found
in the results from the semi-analytic and numerical approaches. On the
whole, however, it is encouraging that two such widely different
treatments lead to essentially the same results.

\subsection{Carbon Production} 
\label{sec:mesacarbon} 

The discussion thus far has focused on the production of $^8$Be
through the process of helium burning. The semi-analytic model
demonstrates that $^8$Be is readily produced. The numerical
simulations not only confirm this result, but also show that helium
burning naturally follows (or partially overlaps with) the
main-sequence phase of hydrogen burning. Nonetheless, the overall goal
of this work is to demonstrate that alternate universes with stable
$^8$Be can produce carbon (as well as oxygen and other heavy elements
necessary for life) without the triple alpha reaction. Universes of
this class will contain $^8$Be from helium burning (as shown here) and
additional $^4$He, both from its early epoch of big bang
nucleosynthesis and from hydrogen burning in ordinary stars. The
required nuclear reaction to produce carbon, $^4$He + $^8$Be $\to$
$^{12}$C, can then take place in a variety of settings.

In some stars, a beryllium burning phase can follow the helium burning
phase, so that carbon is produced in the same stellar core at a later
time. In other cases, however, the star could burn all of its helium
into beryllium, so that no alpha particles are left over for carbon
production in that particular star. Sufficiently massive stars can
continue nucleosynthesis, starting with the reaction $^8$Be + $^8$Be
$\to$ $^{16}$O, and continuing to $^{56}$Fe.  In such cases, some
fraction of the $^8$Be will be returned to the interstellar medium
through stellar winds and/or supernova explosions, so that later
generations of stars will be formed with a mix of $^8$Be and $^4$He.
Carbon can then be produced by those later stellar generations.
Similarly, oxygen can be produced through the reaction $^{4}$He +
$^{12}$C $\to$ $^{16}$O, along with the new possible reaction $^8$Be +
$^8$Be $\to$ $^{16}$O.

As stellar evolution proceeds, the required networks of nuclear
reactions become increasingly complicated \cite{kippenhahn}.  For the
case of helium burning (to produce $^8$Be) considered thus far, we had
to introduce only a single additional composite parameter $\conlum$
that incorporates the cross section for the (single) reaction as well
as the energetic yield (which is determined by the binding energy).
With only one parameter to consider, we could study a range of its
values spanning many orders of magnitude, and could consider a range
of possible stellar masses.  In order to follow the nuclear reaction
chains up the periodic table, however, we would have to introduce
additional nuclear parameters for each possible reaction. The allowed
parameter space for the resulting nuclear reaction network is enormous
and a complete study is beyond the scope of this present paper.  Here
we expand the models to allow for carbon production (see below), which
requires specification of a second nuclear parameter.  However, this
set of simulations does not include nuclear reactions that produce
oxygen, neon, and heavier elements, so the simulations are stopped
once the carbon abundance in the core reaches 50\%.

\begin{figure}
\begin{center}
\includegraphics[scale=0.60]{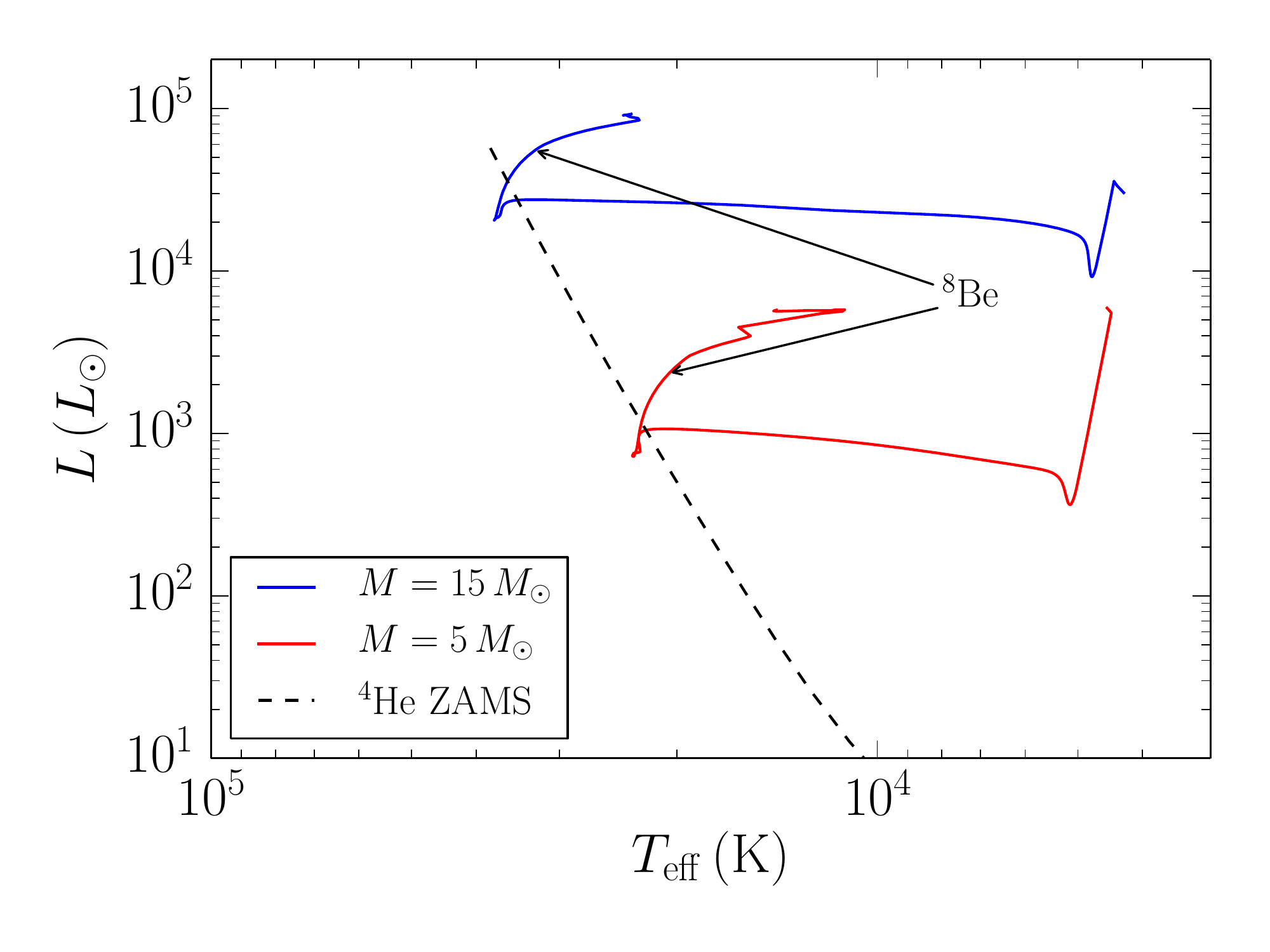}
\end{center}
\caption{Tracks in the H-R diagram for stars in universes with stable  
$^8$Be. The upper blue track corresponds to stellar mass $M_\ast$ = 15
$M_\odot$ and the lower red track corresponds to $M_\ast=5M_\odot$.
The end points of the tracks correspond to the locations in the
diagram where the carbon abundance has reached 50\% in the stellar
core.  The label marks the point along the track where the isotope
$^8$Be reaches a mass fraction of 50\%. The dashed curve marks the
location of the `zero-age' helium burning main-sequence, corresponding
to the point where beryllium production begins.  These simulations
demonstate that carbon can be produced in other universes without the
triple alpha process. }
\label{fig:carbon} 
\end{figure}

We demonstrate the feasibility of carbon production through a
representative set of numerical simulations. To start, we allow the
beryllium and carbon producing reactions to have different rates. For
the helium burning reaction $^4$He + $^4$He $\to$ $^8$Be we use the
nuclear parameter $\conlum=10^{20}$ in cgs units, whereas for the
carbon production reaction $^4$He + $^8$Be $\to$ $^{12}$C we use the
larger value $\conlum_{C}$ = $10^{28}$ in the same units.  Note that
these two nuclear parameters must have different values -- otherwise
the stellar core will burn all of the available helium into beryllium
before any carbon can be produced. For these particular nuclear
parameters, the resulting tracks in the H-R diagram are shown in
Figure \ref{fig:carbon} for two choices of stellar mass, $M_\ast$ = 5
and 15 $M_\odot$. For the sake of definiteness, the stars have
metallicity $Z=10^{-4}$ at the start of their evolution.  Both stars
evolve to a configurations where they burn hydrogen, but relatively
soon enter into a helium burning phase that produces stable $^8$Be
nuclei. Somewhat later, carbon production begins through the burning
of beryllium, while helium burning continues. The simulations are
stopped after the carbon abundance in the stellar core (which
encompasses 10\% of the stellar mass) has reached a mass fraction of
50\%. Note that the tracks also include the pre-main-sequence phases
for both stars. For the $M_\ast=15M_\odot$ star, however, the
pre-main-sequence time scale is shorter than the expected formation
time, so that star is not expected to be optically visible during that
phase.

In the H-R diagram of Figure \ref{fig:carbon}, the dashed curve
depicts an effective zero-age helium burning main-sequence, i.e., the
locus of points where the stars can first burn helium.  This curve is
defined as the epoch when the $^8$Be abundance reaches 2\% in the
stellar core. Note that the onset of hydrogen burning takes place
earlier, when the stars are somewhat dimmer but somewhat hotter. The
tracks are labeled at the benchmark epoch where the abundances of
$^4$He and $^8$Be are both equal to 50\% in the stellar core. At this
epoch, helium burning is well-developed and the stars lie above the
helium-burning zero-age main-sequence marked by the dashed curve.  The
production of carbon begins after the stars evolve further off of this
main sequence.  The end-point of the tracks corresponds to the epoch
when the abundance of $^{12}$C reaches 50\% in the stellar core.

Unlike stars in our universe, where hydrogen fusion dominates the
nuclear reactions for an extended span of time, these stars begin to
burn helium into beryllium soon after the ignition of hydrogen. In our
universe, after stars exhaust the supply of hydrogen in their cores,
they experience significant readjustment in order to burn helium.  An
extreme change in the stellar configuration is necessary to burn
helium through the triple alpha process, which requires a high central
temperature (e.g., see \cite{clayton,kippenhahn,hansen,phil}). In
these stars, however, the required central temperatures are lower (see
Figure \ref{fig:tcm-mesa}) and the adjustment is more modest. As a
result, the stars do not have a clean delineation between their
hydrogen burning phase and their helium burning phase.  Instead, the
two classes of nuclear reactions can take place simultaneously.
Similarly, with the values of the nuclear burning parameters used
here, the synthesis of carbon takes place while the stars continue to
burn helium into beryllium. As a result, during carbon production the
stars are still relatively close to the main-sequence (compared to
stars in our universe). In general, these stars thus have smoother 
transitions between their different burning phases.  

\begin{figure}
\begin{center}
\includegraphics[scale=0.60]{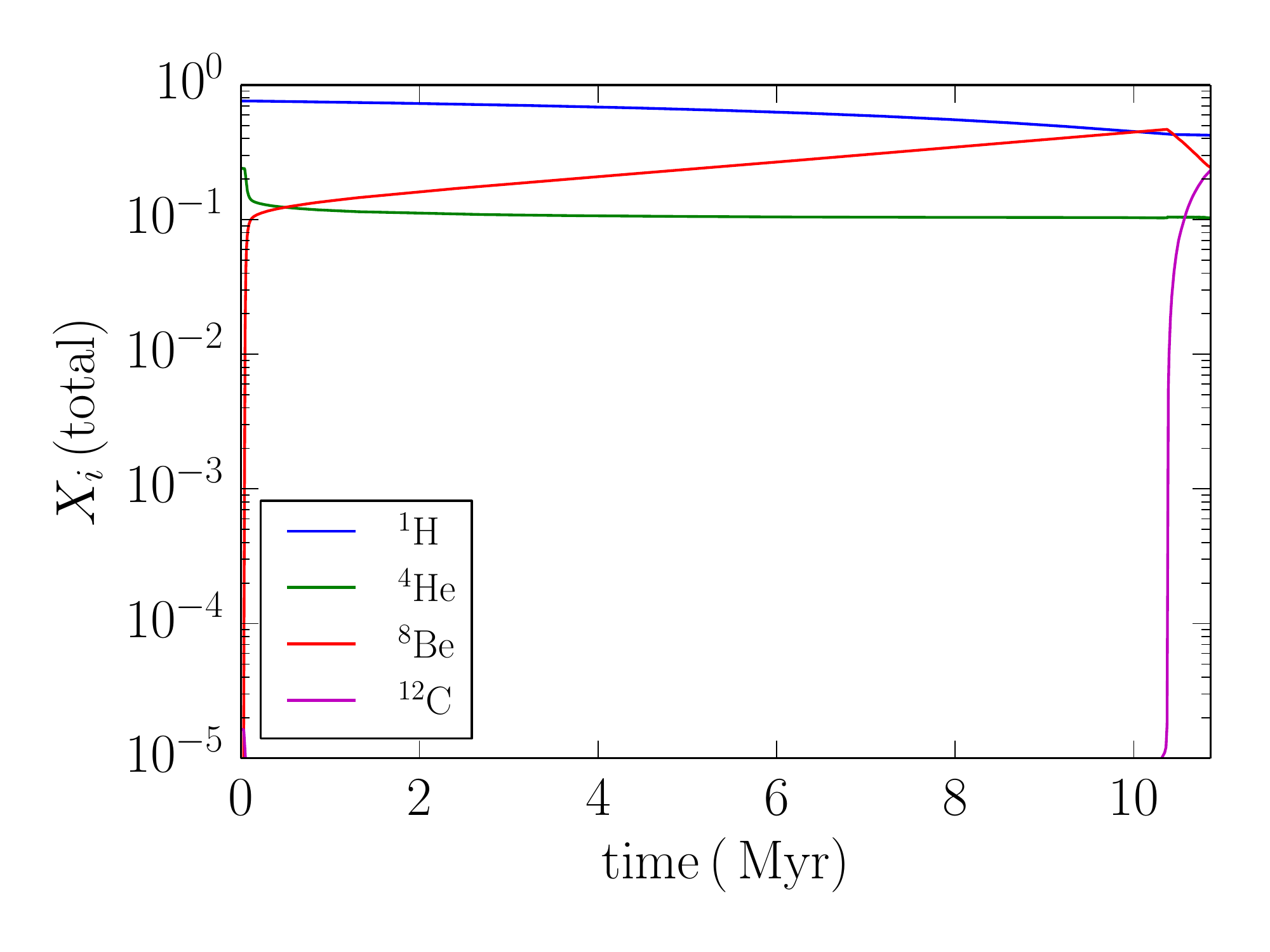}
\end{center}
\caption{Elemental abundances as a function of time for the evolution 
of a star with mass $M_\ast$ = 15 $M_\odot$. The curves show the mass
fractions, averaged over the entire star, versus time for a star with
initial metallicity $Z=10^{-4}$. The nuclear parameters are
$\conlum=10^{20}$ for helium burning and $\conlum_C=10^{28}$ for
beryllium burning (in cgs units). }
\label{fig:x2028m15} 
\end{figure}

Another way to illustrate nuclear processing inside stars is to plot
the abundances of the elements as a function of time. Figure
\ref{fig:x2028m15} shows the mass fractions of hydrogen, helium,
beryllium, and carbon as a function of time for a star with mass
$M_\ast=15M_\odot$. The star starts with metallicity $Z=10^{-4}$ and
thus has essentially no beryllium or carbon at $t=0$. In this star,
the central core is hot enough, and the nuclear burning parameter
$\conlum$ is large enough, that helium burning (beryllium production)
takes place shortly after the star reaches the main-sequence and begins
burning hydrogen. The two nuclear processes take place simultaneously,
so that the mass fraction of $^8$Be increases while the mass fraction
of $^1$H decreases.  The mass fraction of $^4$He initially decreases,
but then reaches a steady state where helium production from hydrogen
burning is approximately balanced by helium burning into beryllium. 
After $\sim10$ Myr, the abundance of beryllium becomes comparable to 
that of hydrogen and the star starts to produce carbon. The simulation
is stopped after 11 Myr because the code requires more complex nuclear
reactions and hence the specification of more $\conlum$-values.

\begin{figure}
\begin{center}
\includegraphics[scale=0.60]{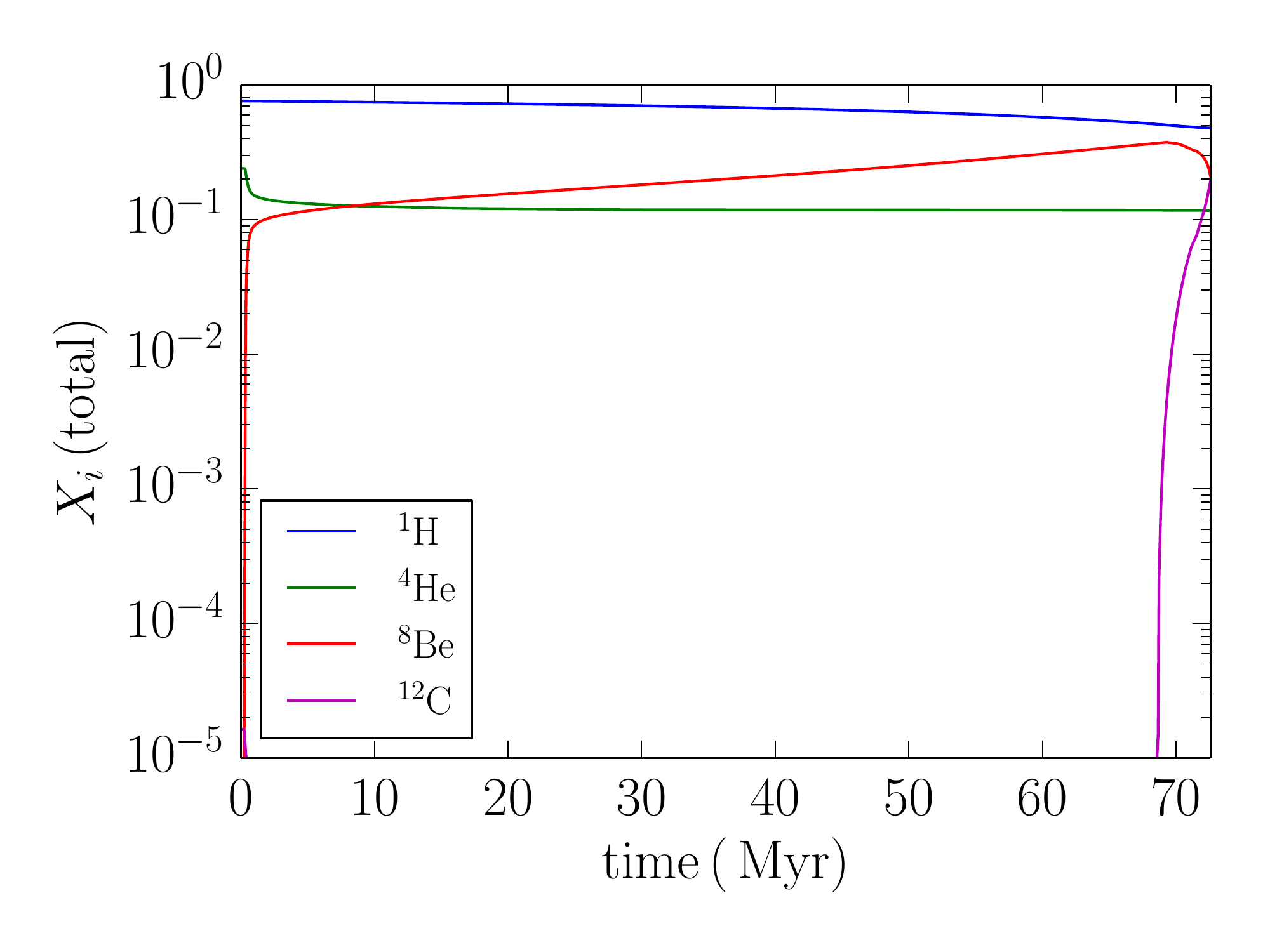}
\end{center}
\caption{Elemental abundances as a function of time for the evolution 
of a star with mass $M_\ast$ = 5 $M_\odot$. The curves show the mass
fractions, averaged over the entire star, versus time for a star with
initial metallicity $Z=10^{-4}$. The nuclear parameters are
$\conlum=10^{20}$ for helium burning and $\conlum_C=10^{28}$ for
beryllium burning (in cgs units). }
\label{fig:x2028m5} 
\end{figure}

Figure \ref{fig:x2028m5} shows the corresponding chemical evolution
diagram for a star with mass $M_\ast=5M_\odot$. The star follows a
similar time sequence. Hydrogen burning takes place afer a brief
pre-main-sequence phase, but the onset of helium burning occurs
shorter thereafter. The two classes of reactions subsequently power
the star for 60 -- 70 Myr. During this epoch, the hydrogen abundance
decreases and the beryllium abundance increases, while helium reaches
a steady-state mass fraction of about 10\%. After $\sim68$ Myr, the
central core grows hot enough to ignite carbon, which increases its
abundance to make up half of the stellar core a few Myr later. Note
that the chemical evolution of the 5 and 15 $M_\odot$ stars are nearly
the same expect for the longer time scale of the smaller star.

For completeness we note that for sufficiently small changes to the
nuclear physics, the triple alpha process should not change, and
carbon production would proceed in the same manner as in our universe
(provided that the Hoyle resonance is present). It is useful to
estimate how much change in the nuclear properties is necessary for
the carbon production scenario to be different. In our universe, a
population of $^8$Be is built up in the stellar core, even though the
isotope is unstable.  In nuclear statistical equilibrium, this
abundance is given by the solution to the nuclear version of the Saha
equation \cite{clayton}, which implies that the mass fraction $X$ of
unstable $^8$Be has the form 
\be
X(^8{\rm Be}) = 1.79 \times 10^{-11} \rho X_\alpha^2 T_9^{-3/2} 
\exp[-1.066/T_9]\,,
\ee
where the quantity in the exponent is given by $1.066/T_9$ = 
$\Delta E_b/kT$.  In this expression, $\Delta E_b=B_8-2B_4=+92$ keV,
where the plus sign indicates that the $^8$Be isotope is unbound.
Under typical conditions in the cores of helium burning stars, where
$\rho\approx10^3$ g cm$^{-3}$, $T_9\approx0.2$, and $X_\alpha\approx1$,
we find the mass fraction $X(^8{\rm Be}) \approx 10^{-9}$. Under these
conditions, the decay mode of $^8$Be is the same as its production
mode, so that the reaction $\alpha + \alpha \leftrightarrow$ $^8$Be
reaches equilibrium.  As long as $X(^8{\rm Be}) \ll 1$, the next step
of the nuclear reaction chain proceeds at a rate such that the mass
fraction $X(^8{\rm Be})$ cancels out \cite{clayton,kippenhahn}. In
order to get a different scenario for carbon production, we thus need
to change the properties of the $^8$Be nucleus so that the equilibrium
abundance predicted by the Saha equation is large enough that the
stellar core can build up its beryllium abundance (instead of
maintaining the trace amounts consistent with nuclear statistical
equilibrium).  For the sake of definiteness, we find the conditions
necessary for an abundance of $X(^8{\rm Be})$ = 0.10, which implies
the constraint 
\be
\exp[(\Delta E_b^{\rm old} - \Delta E_b^{\rm new})/T_9] \sim 10^8 
\qquad {\rm or} \qquad -\Delta E_b^{\rm new} \sim 110 {\rm keV} 
\left( {T \over 10^8 {\rm K}} \right) \,. 
\ee
If stars burning helium in other universes had the same central
temperature as those in our universe, a binding energy of about 200
keV would be sufficient to lead to a different scenario for carbon
production. As shown here, however, the central temperatures can be
lower by factors of 3 -- 10 (see Figure \ref{fig:tcm-mesa}), so that 
a binding energy of only $20-60$ keV can be enough to make the 
equilibrium abundance of $^8$Be large enough to allow for new
scenarios for carbon nucleosynthesis. Note that this energy is
comparable to the energy increment (92 keV) by which $^8$Be is unbound
in our univese and the increment ${\cal O}$(100 keV) by which the
triple alpha resonance can be moved without compromising carbon
production \cite{livio}.

\section{Conclusion} 
\label{sec:conclude} 

This paper argues that the sensitivity of our universe --- and others
--- to the triple alpha reaction for carbon production is more subtle
and less confining than previously reported. If nuclear structures are
different in other universes, so that the carbon resonance is not
present at the proper energy level, then previous work has shown that
carbon production {\it through this process} can be highly suppressed.
If the nuclear structures are different, however, then it remains
possible for the isotope $^8$Be to be stable in other universes. In
such universes with long-lived $^8$Be, carbon production can proceed
without the need for the triple alpha process, or the need for any
particular resonance.

More specifically, the isotope $^8$Be can be stable if its binding
energy is changed by an increment $\sim100$ keV. For comparison, a
larger change in the location of the triple alpha resonance ($\sim300$
keV) is required to compromise carbon production through the standard
reaction network \cite{livio}. In order to produce changes to the
nuclear binding energies of this order, the fundamental constants must
be varied by a few percent (see Section \ref{sec:nuke} and references
therein). Moreover, effective field theory calculations \cite{epelbaum} 
show that the binding energies of $^8$Be and $^4$He do not generally
change at the same rate as the fundamental parameters are
varied,\footnote{More specifically, the derivative of the $^4$He
  binding energy with respect to the fundamental parameters of the
  theory is not equal to half of the derivative of the $^8$Be binding
  energy with respect to the same parameter. This lack of equality is
  necessary to allow changes in the binding energy of $^8$Be relative
  to that of two alpha particles.} so that bound states of $^8$Be can
be realized.

This paper has also shown that, given a stable $^8$Be isotope, stars
can readily produce beryllium through helium burning. We have
addressed this issue using both a semi-analytic model and a
state-of-the-art numerical stellar evolution code ({\sl\small MESA}).
Both approaches are in good agreement and show that the helium burning
main-sequence in other universes corresponds to stars with properties
(luminosities, surface temperatures, central temperatures) that are
roughly similar to those of helium burning stars in our universe. To
carry out these calculations, we have to specify the values of the
nuclear reaction cross sections and the yields for helium burning into
beryllium; in this treatment, these quantities are encapsulated into
the nuclear burning parameter $\conlum$, which we allow to vary by
many orders of magnitude. In spite of the large range of possible
values for $\conlum$, the helium burning main-sequences are
well-defined and vary by less than an order of magnitude in surface
temperature and a factor of $\sim100$ in luminosity (Figures
\ref{fig:hrdiagram} and \ref{fig:hr-mesa}).  Similarly, the central
temperatures vary by about one order of magnitude for the entire range
of $\conlum$ considered here (Figures \ref{fig:centemp} and
\ref{fig:tcm-mesa}). The mass versus luminosity relationship varies by
even less and (approximately) follows the expected scaling law
$L_\ast\propto M_\ast^3$, similar to that for hydrogen burning stars
in our universe (see Figures \ref{fig:masslum} and
\ref{fig:lm-mesa}). It is significant the simplest possible model that
includes the relevant physics (Section \ref{sec:models} and
\cite{adams}) and one of the most sophisticated stellar evolution
codes that is currently available (Section \ref{sec:mesa} and
\cite{paxton}) both give essentially the same results.

After a star produces $^8$Be via helium burning, carbon production can
take place during the subsequent evolution of the same stellar core
and/or during subsequent stellar generations. Using a straightforward
extension of the nuclear reaction network to include $^4$He + $^8$Be
$\to$ $^{12}$C, we have used the {\sl\small MESA} stellar evolution
code to demonstrate that carbon can be produced in stellar cores after
helium burning is sufficiently advanced (see Figures \ref{fig:carbon}
-- \ref{fig:x2028m5}).  This set of simulations thus indicates that
stars can synthesize carbon without the triple alpha process. However,
the abundance of carbon produced in an alternate universe, and the
abundances of other relevant nuclei such as oxygen, depends on a
number of additional properties. In order to follow the chain of
nuclear reactions to ever higher atomic numbers, we would need to
specify the rates and yields for each reaction involving the new
stable isotope $^8$Be, as well as any additional reactions that
operate in that universe (see the discussion below). The ratios of the
various nuclear burning parameters $\conlum$ (defined in equation
[\ref{conlumdef}]) are particularly important. A full assessment of
the possible isotopic ratios in alternate universes is beyond the
scope of any single paper and is left for future work.

We note that the apparent fine-tuning problem posed by the triple
alpha process has (at least) two components. One can ask if the
$^{12}$C resonance is necessary to produce the abundances of carbon,
oxygen, and other heavy elements observed in our universe. On the
other hand, one can consider what nuclear properties are required to
produce enough carbon for some type of life to exist. Previous work
has shown that nuclear physics cannot be changed greatly without
altering the abundances of carbon and oxygen observed in our universe,
which utilizes the triple alpha process.  In contrast, this paper
shows that the triple alpha process is not a necessary ingredient for
carbon production in other universes.

The results of this paper do not predict the range of possible isotope
ratios. Such a determination must consider the rates (and yields) for
all of the relevant nuclear reactions, where these rates can vary from
universe to universe. A related complication is that the reaction
rates are secondary parameters that are ultimately determined by the
underlying fundamental theory. The changes in the reaction rates and
other nuclear properties are thus coupled, but the coupling is not
known. This paper has focused on non-resonant two-body reactions, and
shows that carbon can be produced without the triple alpha resonance.
However, resonances are simply manifestations of the energy levels of
the interacting nuclei, so that every universe is expected to have
some collection of resonances, which will in turn affect the isotopic
ratios. The locations of these resonances can favor some nuclear
reactions over others (by enhancing particular cross sections) and
will help produce a wide range of different isotope abundances across
the multiverse.

In the absence of a specification of the possible reaction rates,
yields, and resonances, one can explore the parameter space for all of
the possible nuclear reactions (e.g., with and without stable $^8$Be)
and determine which cases produce favorable isotopic ratios. In
addition to the large possible parameter space, such a program is
further complicated because stars of varying masses produce different
nuclear yields, and the distribution of stellar masses can also vary.
Moreover, even if the allowed parameter space for producing given
abundances of carbon could be fully delineated, we do not know how
much carbon is actually required for life. Finally, the probability of
a universe being habitable ultimately depends not only on the range of
allowed parameters, but also on the distribution from which the
parameters are sampled; unfortunately, this underlying probability
distribution also remains unknown. As a result of these complications,
this paper represents only one step toward understanding carbon
production in other universes (see also
\cite{carr,bartip,hogan,aguirre,barnes2012,schellekens} and references
therein).

Although carbon is (most likely) necessary for a universe to be
habitable, it is not sufficient. Universes favorable for the
development of life require additional heavy elements, including
oxygen, nitrogen, and many others. Although a full treatment of heavy
element production for all possible universes is beyond the scope of
this paper, we can outline some basic requirements. Hydrogen is a
necessary ingredient, so it is important that big bang nucleosynthesis
does not process all of the protons into heavier nuclei and that star
formation is not overly efficient.  The natural endpoint for stellar
nucleosynthesis is to produce large quantities of the element with the
highest binding energy per particle, i.e., iron (in our universe) or
its analog (in other univeres).  As a result, nuclear processing in
stars cannot be too efficient. In our universe, stars span a range of
masses, from those that can barely burn hydrogen up to those that
produce iron cores and explode as supernovae. This range of stellar
masses results in a wide range of endpoints for nuclear reaction
chains and is thus favorable for producing a diverse ensemble of heavy
elements. Habitable universes thus reside in a regime with
intermediate properties. Star formation must take place readily in
order to produce the heavy elements, energy, and planets necessary for
life, but cannot be so efficient that no hydrogen is left over for
water.  Stars must be able to synthesize the full distribution of
heavy elements necessary for life, but cannot be so efficient that all
nuclei become iron (or whatever nuclide has the highest binding energy
in the given universe). This paper generalizes the class of
potentially habitable universes to include those without the triple
alpha process for carbon production, but a great deal of additional
work remains to sort out the full parameter space.

\bigskip 
\noindent
{\bf Acknowledgments:} We would like to thank Anthony Aguirre, Kate
Coppess, Juliette Becker, George Fuller, Minhyun Kay, Mark Paris,
Nicole Vassh, Stan Woosley, and Coco Zhang for useful discussions. 
We also thank the referee for a prompt and inciteful report. This 
work was supported by the JT Foundation through Grant ID55112: 
{\sl Astrophysical Structures in other Universes}, and by the
University of Michigan.


\end{document}